\documentclass[11pt,letter,subeqn]{article}
\usepackage{amsmath, amssymb,amsthm,cite}
\usepackage{graphics,color}
\usepackage{graphicx}
\usepackage{tabulary}
\usepackage{float}
\usepackage{caption, subcaption}
\usepackage{subfiles}
\usepackage{multirow}
\usepackage{chngcntr}
\usepackage{cleveref}
\counterwithin{figure}{section}
\counterwithin{table}{section}

\title{On Exact and Approximate Symmetries of Algebraic and Ordinary Differential Equations with a Small Parameter }

\author{ 
Mahmood R. Tarayrah\footnotemark[1],~~ Alexei F. Cheviakov \footnotemark[2]\vspace{0.5cm}\\
\small $^{\rm a,b}$\emph{Department of Mathematics and Statistics, University of Saskatchewan, Saskatoon, Canada}\vspace{0.2cm}\\
}

\def\beq{\begin{equation}}
\def\eeq{\end{equation}}
\def\barr{\begin{array}{ll}}
\def\earr{\end{array}}

\def\const{\hbox{\rm const}}

\newtheorem{theorem}{Theorem}[section]

\newtheorem{proposition}{Proposition}

{\theoremstyle{definition} 
\newtheorem{example}{Example}[section]
\newtheorem{remark}{Remark}[section]
}

\newcounter{tabnum}

\begin{document}


\footnotetext[1]{Corresponding author. Electronic mail: mrt566@usask.ca}
\footnotetext[2]{Alternative English spelling: Alexey Shevyakov. Electronic mail: shevyakov@math.usask.ca}

\maketitle \numberwithin{equation}{section}
\maketitle \numberwithin{remark}{section}
\numberwithin{lemma}{section}
\numberwithin{proposition}{section}

\begin{abstract}

The framework of Baikov-Gazizov-Ibragimov approximate symmetries has proven useful for many examples where a small perturbation of an ordinary differential equation (ODE) destroys its local symmetry group. For the perturbed model, some of the local symmetries of the unperturbed equation may (or may not) re-appear as approximate symmetries, and new approximate symmetries can appear. Approximate symmetries are useful as a tool for the construction of approximate solutions. We show that for algebraic and first-order differential equations, to every point symmetry of the unperturbed equation, there corresponds an approximate point symmetry of the perturbed equation. For second and higher-order ODEs, this is not the case: some point symmetries of the original ODE may be unstable, that is, they do not arise in the approximate point symmetry classification of the perturbed ODE. We show that such unstable point symmetries correspond to higher-order approximate symmetries of the perturbed ODE, and can be systematically computed. Two detailed examples, including a fourth-order nonlinear Boussinesq equation, are presented. Examples of the use of higher-order approximate symmetries and approximate integrating factors to obtain approximate solutions of higher-order ODEs are provided.

\end{abstract}


\section{Introduction}

A symmetry of a system of algebraic or differential equations is a transformation that maps solutions of the system to other solutions. Dating back to the works of Sophus Lie in the nineteenth century, symmetry ideas have seen significant development over the last century, relating to symmetry reduction and solution of differential equations, integrating factors, conserved quantities and local conservation laws, Hamiltonian and Lagrangian structure, integrability, nonlocal extensions, invertible and non-invertible mappings between different classes of differential equations, and more (see, e.g., Refs.~\cite{olver2000applications, bluman2010applications} and references therein).

Let the term \emph{perturbed equations} denote equations differing from some canonical or otherwise well understood model by extra term(s) with involving a small parameter. This small perturbation disturbs local Lie symmetry properties of the unperturbed equations.
Several approximate Lie symmetry methods have been developed to study symmetry properties of perturbed models, and relate and compare them to symmetry structure of the unperturbed equations. The approximate symmetry method (referred to here as the \emph{BGI method}) was introduced by Baikov, Gazizov and Ibragimov \cite{baikov1989, baikov1991, baikov1993}, where the approximate symmetry generator is expanded in a perturbation series. This approach preserves the Lie group structure, in particular, a commutator of two approximate symmetry generators yields an approximate symmetry generator \cite{gazizov1996lie}. Using the BGI framework, approximate symmetries, first integrals, and approximate solutions have been constructed for a number of models involving ordinary and partial differential equations (ODE, PDE) (e.g., Refs.~\cite{unal2000periodic, baikovjoining, bai2018approximate}.) A different approach to approximate symmetries, developed by Fushchich and Shtelen \cite{fushchich1989approximate}, combines a perturbation technique with the symmetry group method by expanding the dependent variables in a Taylor series in the small parameter, and approximately replacing the original equations by a system of equations that are coefficients at different powers of the parameter. The classical Lie symmetry method is applied to obtain symmetries of the latter system. Using this method, approximate symmetries and approximate solutions have been found for some PDE models \cite{euler1992approximate,euler1994}. The BGI and Fushchich-Shtelen approaches are not equivalent. They have been compared and used to obtain approximate symmetries and approximate solutions for several PDE models \cite{wiltshire2006two,grebenev2007approximate, mahdavi2015two,ahmed2017invariant}.

In Ref.\cite{jiao2008approximate}, approximate solutions of a singularly perturbed Boussinesq PDE were found using approximate Fushchich-Shtelen symmetries and a specialized perturbation method. Burde \cite{burde2001potential} developed a new approach for approximate symmetries by constructing equations that could be reduced by exact transformations to an unperturbed equation and at the same time would coincide approximately with the perturbed equation.

Contact and higher-order exact symmetries can be used to construct solutions for ordinary differential equations (e.g., \cite{bluman2008symmetry}). In Ref.~\cite{ibragimov2006integrating}, it was shown how integrating factors for linear and nonlinear ordinary differential equations can be determined. A perturbation method based on integrating factors was developed for a system of regularly perturbed first-order ODEs \cite{van1999perturbation}.

In this paper, we follow the BGI approximate symmetry framework for algebraic equations and ODEs. While BGI approximate symmetries have been found for many models, including ODEs and PDEs (e.g., \cite{grebenev2007approximate,unal2000periodic,baikov1989,baikov1991,baikov1993,ibragimov1995crc,gazizov1996lie}), it has not been clarified under what conditions point or local symmetries of exact equations can become unstable, disappearing from the classification of approximate symmetries of a perturbed system of the same differential order, and what form they therefore take. In the sections below, we prove that for single algebraic and first-order differential equations, all point symmetries are stable. For second and higher-order ODEs, we show that a point or local symmetry of the unperturbed equation usually yields a higher-order (generally, of order $n-1$) symmetry of the perturbed model.

The paper is organized as follows. In Section \ref{sec:PAs}, we briefly review the necessary notation for the Lie group of transformations, infinitesimal transformations and determining equations for finding the exact symmetries, and provide an introduction to the theory of approximate
transformations and approximate symmetries of the perturbed equations in the sense of Baikov, Gazizov and Ibragimov. In Section \ref{sec:Alg}, we consider exact and approximate point symmetries of algebraic equations with a small parameter, and provide the relation between exact and approximate symmetries of the original and perturbed algebraic equations. In Section \ref{sec:1stODEs}, we study the exact and BGI approximate symmetries of perturbed first-order ordinary differential equations. In Section \ref{sec:higherODEs}, we investigate the BGI approximate symmetries of the perturbed higher-order ODEs. It is shown that point symmetries of the unperturbed equation may indeed disappear from the classification of approximate point symmetries of the perturbed model, and conditions for that are given. In Section \ref{sec:localsym}, we consider point and higher-order local exact and approximate symmetries of second and higher-order ODEs in evolutionary form, and present a systematic way to find approximate symmetry components for approximate symmetries that correspond to \emph{every} point and local symmetry of the unperturbed equation. Relations between exact and approximate symmetries are considered in detail for two examples, including a nonlinearly perturbed second-order ODE, and a fourth-order ODE arising as a traveling wave reduction of the Boussinesq partial differential equation modeling  shallow water wave propagation. Finally, in Section \ref{sec:sols}, we determine the approximate integrating factors of perturbed first-order ODEs using approximate point symmetries. We find the determining equations of approximate integrating factors, and show how these determining equations and higher-order approximate symmetries can be used to obtain approximate solutions of a perturbed Boussinesq ODE. A brief discussion is offered in Section \ref{conclusion}.

\section{Lie groups of exact and approximate point and local symmetries}\label{sec:PAs}

We now briefly overview the framework of Lie point and local symmetries, in comparison with the BGI approximate symmetry framework \cite{baikov1989, baikov1991, baikov1993} for models involving a small parameter.

\subsection{Local symmetries of algebraic and differential equations}

A general system of $N$ algebraic or differential equations is given by
\begin{equation}\label{gs1}
   F^{\sigma}(x,v,\partial v,\ldots, \partial^k v)=0,\quad k\geq 0, \quad \sigma=1,2,...,N.
\end{equation}
Here $x = (x^1 ,x^2 ,...,x^n )$, $n\geq 1$, and $v=(v^1, v^2, ..., v^m)$, $m\geq 1$, denote respectively independent and dependent variables, and a symbol $\partial^q v$ denotes all $q^{\rm th}$-order derivatives of all components of $v$.
A one-parameter Lie group of transformations in the space of the problem variables $(x,v)$ is given by
\begin{equation}\label{e25}
\begin{array}{l}
  (x^*)^i  = f^i (x,v;a) =x^i+a\xi^i(x,v)+ O(a^2) ,\quad i = 1,2, \ldots ,n, \\
  (v^*)^\mu  = g^\mu (x,v;a)= v^\mu +a \eta^\mu(x,v)+ O(a^2),\quad \mu = 1,2, \ldots ,m, \\
 \end{array}
\end{equation}
with the group parameter $a$, and the corresponding infinitesimal generator
\begin{equation}\label{e100}
X = \xi^i (x,v)\dfrac{\partial }{{\partial x^i }}
+ \eta ^\mu (x,v) \dfrac{\partial }{{\partial v^\mu }},
\end{equation}
where (as well as below, where appropriate) summation in repeated indices is assumed, and
\[
\xi^i(x,v) = \dfrac{\partial f^i}{\partial a},\qquad \eta^\mu (x,v) = \dfrac{\partial g^\mu }{\partial a}
\]
denote the infinitesimal components (see, e.g., \cite{olver2000applications, bluman2010applications} and references therein).
One can reconstruct the global Lie group \eqref{e25} from its infinitesimals using the initial-value problem
\begin{equation}\label{Lie1}
\barr
  \dfrac{d (x^*)^i}{d a} = \xi^i(x^*,v^*),\quad (x^*)^i\big|_{a=0}=x^i, \\[1.5ex]
  \dfrac{d (v^*)^\mu}{d a} = \eta^\mu(x^*,v^*),\quad (v^*)^\mu\big|_{a=0}=v^\mu.
\earr
\end{equation}
The $k^{\rm th}$  prolongation of the Lie group of transformations \eqref{e25} transforms the derivatives as follows:
\begin{equation}\label{e26}
\begin{array}{l}
  (v^*)^\mu_i  = v_i^\mu  + a \eta _i^{(1)\mu}(x,v,\partial v) + O(a^2), \\
  \vdots  \\
 (v^*)_{i_1 i_2  \ldots i_k }^\mu  = v_{i_1 i_2  \ldots i_k }^\mu  + a \eta _{i_1 i_2  \ldots i_k }^{(k)\mu}(x,v,\partial v,\ldots, \partial^k v) + O(a^2) \\
 \end{array}
\end{equation}
acting on the $(x,v,\partial v,\ldots,\partial^k v)$ jet space. The extended infinitesimals $\eta _{i_1 \ldots i_k }^{(k)\mu}$ appearing above are given by the  prolongation formulas
\begin{equation}\label{ext inf}
 \eta _i^{(1)\mu}  = D_i \eta ^\mu  - v_j^\mu D_i \xi^j , \qquad \eta _{i_1 i_2  \ldots i_k }^{(s)\mu}  = D_{i_s} \eta _{i_1 i_2  \ldots i_{s - 1} }^{(s - 1)\mu}  - v_{i_1 i_2  \ldots i_{s - 1} j}^\mu D_{i_s } \xi^j ,
\end{equation}
 $\mu = 1,2, \ldots ,m,$ $i,i_j  = 1,2, \ldots ,n$ for \,$s = 1,2, \ldots ,k$. $D_i$ is the total derivative operator given by
\begin{equation}\label{DO}
D_i  = \dfrac{\partial }{{\partial x^i }} + v_i^\mu \dfrac{\partial
}{{\partial v^\mu }} + v_{ij}^\mu \dfrac{\partial }{{\partial v_j^\mu }}
+\ldots + v_{ii_1 i_2  \ldots i_n }^\mu \dfrac{\partial
}{{\partial v_{i_1 i_2  \ldots i_n }^\mu }} +  \ldots .
\end{equation}
The $k^{\rm th}$ prolongation of the infinitesimal generator \eqref{e100} is
 \begin{equation}\label{e101}
   X^{(k)} = \xi^i \dfrac{\partial }{{\partial x^i }}
+ \eta ^\mu \dfrac{\partial }{{\partial v^\mu }}+\eta _i^{(1)\mu}\dfrac{\partial}{\partial  v_i^\mu }+\ldots +\eta _{i_1 i_2  \ldots i_k }^{(k)\mu}\dfrac{\partial}{\partial
v^\mu_{i_1 i_2  \ldots i_k}}, \quad k\geq 1.
 \end{equation}
The one parameter Lie group of transformations \eqref{e25} is a point symmetry of the system \eqref{gs1} when for each
$\sigma=1,2,...,N,$
\begin{equation}\label{det:eq:exact}
X^{(k)} F^{\sigma}(x,v,\partial v,\ldots, \partial^k v)=0
\end{equation}
holds on solutions of \eqref{gs1} (e.g., \cite{olver2000applications, bluman2010applications}). The evolutionary (characteristic) form of the Lie group of transformations \eqref{e25} is the one-parameter family of transformations that leaves invariant the
independent variable $x$:
\begin{equation}\label{evol}
\begin{array}{l}
(x^*)^i  =x^i ,\quad i = 1,2, \ldots ,n, \\
(v^*)^\mu  = v^\mu +a [{\eta}^\mu(x,v)-v_i^\mu \xi^i(x,v)]+ O(a^2),\quad \mu = 1,2, \ldots ,m, \\
\end{array}
\end{equation}
with the evolutionary infinitesimal generator
\begin{equation}\label{evolg}
\hat{X}=[{\eta}^\mu(x,v)-v_i^\mu \xi^i(x,v)]\frac{\partial}{\partial v^\mu}.
\end{equation}
Higher-order local transformations generalize \eqref{evol} by allowing the infinitesimal components to depend on higher derivatives of $v$:
\begin{equation}\label{higher}
\begin{array}{l}
  (x^*)^i  =x^i ,\quad i = 1,2, \ldots ,n, \\
  (v^*)^\mu  = v^\mu +a \zeta^\mu(x,v,\partial v,...,\partial^{s} v)+ O(a^2),\quad \mu = 1,2, \ldots ,m, \\
 \end{array}
\end{equation}
where $\zeta^\mu[v] = \zeta^\mu(x,v,\partial v,...,\partial^{s} v)$ are differential functions. The corresponding infinitesimal generator is given by
\begin{equation}\label{highgenerator1}
   \hat{X}=\zeta^\mu[v]\dfrac{\partial}{\partial v^\mu}.
\end{equation}
The prolongation of \eqref{highgenerator1} is defined by
\begin{equation}\label{highprolong}
   \hat{X}^\infty=\zeta^\mu\dfrac{\partial}{\partial v^\mu}+\zeta_{i}^{(1)\mu} \dfrac{\partial }{{\partial v_i^\mu }}
+\ldots + \zeta_{i_1 i_2  \ldots i_p }^{(p)\mu} \dfrac{\partial
}{{\partial v_{i_1 i_2  \ldots i_p }^\mu }} +  \ldots ,
 \end{equation}
where the higher-order components are computed using
\begin{equation}\label{ho:symm:comp}
   \zeta_{i}^{(1)\mu} = D_i\zeta^\mu, \qquad
   \zeta_{i_1 i_2  \ldots i_p }^{(p)\mu} =D_{i_p}\zeta_{i_1 i_2  \ldots i_{p-1} }^{(p-1)\mu},
 \end{equation}
for $\mu=1,\ldots,m$; $i,i_j=1,\ldots,n$, $p=2,3,\ldots$. As it is well-known, transformations \eqref{higher} usually cannot be integrated to yield a global group like \eqref{e25}, because the analog of the ODE system in the Cauchy problem \eqref{Lie1} will contain an infinite countable number of equations, with right-hand sides involving derivatives of $v^*$ of higher orders than the left-hand sides. The only exception occurs, for first-order \emph{contact transformations}, arising in the case of a single dependent variable $v$ ($m=1$) from
generators
\begin{equation}\label{contactX}
  \hat{X}=\zeta(x,v,\partial v)\dfrac{\partial}{\partial v}.
\end{equation}
In this case, the global group can be found explicitly.

\medskip\begin{example}\label{Example A-unperturbed}
The classical example of a second-order ODE with a maximal Lie group of point symmetries is
\begin{equation}\label{ypp:0}
y''(x)=0.
\end{equation}
Here and below, we use primes to denote derivatives. (The perturbed version of \eqref{ypp:0} will be used to compute examples of approximate symmetries below.) Let $X^0=\xi^0(x,y){\partial}/{\partial x}+\eta^0(x,y) {\partial}/{\partial y}$
denote the point symmetry generator admitted by the ODE \eqref{ypp:0}. The computation of the prolongation of $X^0$ to the second order and solution of the determining equations \eqref{det:eq:exact} yield the general solution (e.g., \cite{olver2000applications})
\beq\label{symm:Ypp0}
\barr
  \xi^0 = C_1x^2+C_3\dfrac{xy}{2}+C_7x+C_6y+C_8,\\
  \eta^0 = C_1xy+C_2 x+C_3\dfrac{y^2}{2}+C_4y+C_5,
\earr
\eeq
where $C_i$ are arbitrary constants. The resulting eight-parameter Lie group of point symmetries of \eqref{ypp:0} is spanned by the generators
\beq\label{symm:Ypp0:generators}
\barr
  X_1^0 = xy\dfrac{\partial}{\partial y}+ x^2\dfrac{\partial}{\partial x},\quad X_2^0=x\dfrac{\partial}{\partial y},\quad X_3^0= \dfrac{y^2}{2}\dfrac{\partial}{\partial
  y}+\dfrac{xy}{2}\dfrac{\partial}{\partial x},  \\[2ex]
  X_4^0 = y\dfrac{\partial}{\partial y},\quad X_5^0=\dfrac{\partial}{\partial y},\quad X_6^0= y\dfrac{\partial}{\partial x},\quad X_7^0= x\dfrac{\partial}{\partial x},\quad
  X_8^0= \dfrac{\partial}{\partial x}.
\earr
\eeq
\end{example}

\subsection{Perturbed equations: exact and approximate symmetries}

\subsection*{\textbf{(A)} Exact symmetries of perturbed equations}

Consider a system of algebraic or differential equations
\begin{equation}\label{unperturbed eq}
  F_0^{\sigma}(x,v,\partial v,\ldots, \partial^k v)=0, \quad  \sigma=1,2,...,N,
\end{equation}
where $k\geq 0$ is the differential order, and its first-order perturbation in terms of a small parameter $\epsilon$ is written as:
\begin{equation}\label{eq:perturbed:general}
\barr
   F^{\sigma}(x,v,\partial v,\ldots, \partial^k v;\epsilon)=F_0^{\sigma}(x,v,\partial v,\ldots, \partial^k v)+\epsilon F_1^{\sigma}(x,v,\partial v,\ldots, \partial^k v)=0, \\[2ex]
   \sigma=1,2,...,N.
\earr
\end{equation}
(The right-hand side of \eqref{eq:perturbed:general} can be replaced by $o(\epsilon)$.) Denote the exact point and local symmetry generators of \eqref{eq:perturbed:general} by
\begin{equation}\label{m14}
  Y=\alpha^{i}(x,v;\epsilon)\dfrac{\partial}{\partial x^i}+ \beta^{\mu}(x,v;\epsilon)\dfrac{\partial}{\partial v^\mu},
\end{equation}
\begin{equation}\label{highgenerator}
   \hat{Y}=\zeta^\mu(x,v,\partial v,...,\partial^{s} v;\epsilon)\dfrac{\partial}{\partial v^\mu}.
\end{equation}
Solving the determining equations \eqref{det:eq:exact}, one finds exact symmetries of \eqref{eq:perturbed:general}, holding for an arbitrary $\epsilon$. It is commonly the case that due to the perturbation term, some or even all point and/or local symmetries of the unperturbed equations \eqref{unperturbed eq} disappear from the local symmetry classification of the perturbed model \eqref{eq:perturbed:general}.
\begin{example}\label{ex:exact-of-perturbed}
  Consider an ODE
\begin{equation}\label{pert eqn1}
  y''=\epsilon (y')^{-1},
\end{equation}
which is a perturbed version of \eqref{ypp:0}. It can be shown that the only symmetries of \eqref{ypp:0} that are also symmetries of \eqref{pert eqn1}, holding for an arbitrary $\epsilon$, are the translations
\begin{equation}\label{}
  Y_1=X_5^0=\dfrac{\partial}{\partial y},\quad Y_2=X_8^0= \dfrac{\partial}{\partial x}.
\end{equation}
\end{example}

\subsection*{\textbf{(B)} Approximate symmetries of perturbed equations}
Approximate symmetries present a tool to seek additional symmetry structure of perturbed models. For equations \eqref{eq:perturbed:general} involving a small parameter $\epsilon$, we now define Baikov-Gazizov-Ibragimov (BGI) approximate point symmetries \cite{ibragimov1995crc,ibragimov2009approximate}. A \textit{one-parameter Lie group $G$ of BGI approximate point transformations} acting on $(x,v)$-space is given by
\begin{equation}\label{approx}
\begin{array}{ll}
     {(x^*)}^{i} ~=~ f^i(x,v;a,\epsilon) ~=~ f_0^i(x,v;a)+\epsilon f_1^i(x,v;a) + o(\epsilon),&  i=1,...,n, \\[1ex]
     {(v^*)}^{\mu}~ =~g^\mu(x,v;a,\epsilon)~=~ g_0^\mu(x,v;a)+\epsilon g_1^\mu(x,v;,a) + o(\epsilon),&  \mu=1,...,m,
\earr
\end{equation}
where $f_j^i$, $g_j^\mu$ are sufficiently smooth functions, and $a$ is the group parameter. The generator of the approximate group $G$ is given by
\beq\label{X:X0:X1}
\barr
X &= X^0+\epsilon X^1 \\[2ex]
  &= \left(\xi_0^i(x,v)+\epsilon \xi_1^i(x,v)\right)\dfrac{\partial}{\partial x^i}+\left(\eta_0^\mu(x,v)+\epsilon \eta_1^\mu(x,v)\right)\dfrac{\partial}{\partial v^\mu},
\earr
\eeq
where
\[
\xi_j^i=\frac{\partial f_j^i(x,v,a) }{\partial a}\bigg|_{a=0},\quad \eta_j^\mu=\frac{\partial g_j^\mu(x,v,a) }{\partial a}\bigg|_{a=0},\quad  j=0,1,\quad i=1,...,n,\quad
\mu=1,...,m
\]
are the infinitesimal components. One can reconstruct the global group of approximate transformations from its generator components using the equations
\beq\label{eq:Lie:approx}
\barr
\dfrac{df_0^i}{da}= \xi_0^i(f_0,g_0), \quad \dfrac{df_1^i}{da}=\displaystyle \sum_{k=1}^{n}\frac{\partial \xi_0^i}{\partial x^k}\bigg|_{(x,v)=(f_0,g_0)}f_1^k + \xi_1^i(f_0,g_0),\\[3ex]
 f_0^i\big|_{a=0}=x^i,\,\quad  f_1^i\big |_{a=0}=0,\, i=1,...,n,\\[3ex]
\dfrac{dg_0^\mu}{da}= \eta_0^\mu(f_0,g_0),\quad  \dfrac{dg_1^\mu}{da}=\displaystyle \sum_{k=1}^{m}\frac{\partial \eta_0^\mu}{\partial v^k}\bigg|_{(x,v)=(f_0,g_0)}f_1^k +\eta_1^\mu(f_0,g_0),\\[3ex]
g_0^\mu\big|_{a=0}=v^\mu,\quad  g_1^\mu\big |_{a=0}=0,\, \mu=1,...,m.
\earr
\eeq

\begin{remark}
Similarly to exact higher-order local transformations \eqref{higher}, \eqref{highgenerator1}, one can define local approximate BGI transformations, with generators in evolutionary form given by
\beq\label{X:X0:X1:higher}
\hat{X} = \hat{X}^0+\epsilon \hat{X}^1 = \left(\zeta_0^\mu[v]+\epsilon \zeta_1^\mu[v]\right)\dfrac{\partial}{\partial v^\mu}.
\eeq
\end{remark}

\begin{remark}
In certain cases, such as for differential equations involving several terms involving different orders of the small parameter $\epsilon$, one can seek approximate symmetries with generators of the form
\[
\barr
  X = \left(\xi_0^i(x,v)+\epsilon \xi_1^i(x,v)+...+\epsilon^p \xi_p^i(x,v) \right)\dfrac{\partial}{\partial x^i}  +\left(\eta_0^\mu(x,v)+\epsilon \eta_1^\mu(x,v)+...+ \epsilon^p \eta_p^\mu(x,v)\right)\dfrac{\partial}{\partial v^\mu}
\earr
\]
for an arbitrary order $p \geq 1$ \cite{ibragimov1995crc}.
\end{remark}
\begin{example}
Let $n=1$, and consider a generator
\beq\label{eg:2:X}
  X=(1+\epsilon x)\dfrac{\partial}{\partial x}.
\eeq
If \eqref{eg:2:X} is treated as a generator of an approximate group \eqref{approx}, with $\xi_0(x)=1$ and $\xi_1(x)=x$, the Lie's equations \eqref{eq:Lie:approx} become
\[
\dfrac{df_0}{da}=1, \quad f_0\big|_{a=0}=x, \quad \dfrac{df_1}{da}=f_0, \quad f_1\big|_{a=0}=0,
\]
with the solution $f_0=x+a$, $f_1=ax+{a^2}/{2}$, leading to the global approximate transformation group
\beq\label{eq:eg2:appr}
  {x^*}= x+a+\epsilon \left(ax+\dfrac{a^2}{2}\right).
\eeq
If \eqref{eg:2:X} is considered as an exact generator of a Lie group, then solving the Lie's equation \eqref{Lie1} yields the global group
\beq\label{eq:eg2:exact}
   {x^*}=xe^{a\epsilon}+\dfrac{e^{a\epsilon}-1}{\epsilon}=x+a+\epsilon \left(ax+\dfrac{a^2}{2}\right)+\epsilon^2\left(\dfrac{a^2}{2}x+\dfrac{a^3}{6}\right)+\ldots\,,
\eeq
where the Taylor expansion of the transformed $x$ in the small parameter contains the approximate group \eqref{eq:eg2:appr} as the first three terms.
\end{example}

\subsection*{\textbf{(C)} Approximate invariance. Computation of BGI approximate symmetries. Trivial approximate symmetries}

Let $G$ be an approximate group of BGI point transformations \eqref{approx}. A system of perturbed algebraic or differential equations \eqref{eq:perturbed:general}
is \textit{approximately invariant} with respect to $G$ if
\[
F^\sigma(x^*,v^*,\partial v^*,\ldots, \partial^k v^*;\epsilon)=o(\epsilon), \quad \sigma=1,...,N,
\]
whenever $F^{\alpha}(x,v,\partial v,\ldots, \partial^k v;\epsilon)=0$ for $\alpha=1,...,N$. The following theorem holds \cite{ibragimov1995crc}.
\begin{theorem}\label{1002}
Let the equations \eqref{eq:perturbed:general} be approximately invariant under the approximate group $G$ of point transformations \eqref{approx} with the generator \eqref{X:X0:X1}
such that $\xi^0, \eta^0(x,v) \neq 0$. Then the infinitesimal operator
\begin{equation}\label{60}
X^0=\xi_0^i(x,v)\dfrac{\partial}{\partial x^i}+ \eta_0^\mu(x,v)\dfrac{\partial}{\partial v^\mu}
\end{equation}
is a generator of an exact symmetry group for the unperturbed equations \eqref{unperturbed eq}.
\end{theorem}
The determining equations to find the components $\xi_j^i$, $\eta_j^\mu$ (for approximate symmetries of the system \eqref{eq:perturbed:general}) are given by
\begin{equation}\label{detapprox}
\barr
    (X^{0^{(k)}}+\epsilon X^{1^{(k)}})(F_0^{\sigma}(x,v,\partial v,\ldots, \partial^k v)+\epsilon F_1^{\sigma}(x,v,\partial v,\ldots, \partial^k v))\bigg |_{F_0+\epsilon
    F_1=0}=o(\epsilon),\\[2ex]
    \sigma=1,...,N.
\earr
\end{equation}
The Hadamard's lemma \cite{olver2000applications} allows the following equivalent reformulation of \eqref{detapprox}.
\begin{theorem}
There exist smooth differential functions $P_{\sigma \tau}(x,v,\partial v,\ldots, \partial^k v)$, $Q_{\sigma \tau}(x,v,\partial v,\ldots, \partial^k v)$ such that
\begin{eqnarray}
X^{0^{(k)}}F_0^\sigma &=& \sum_{\tau=1}^{N} P_{\sigma \tau} F_0^\tau,\label{3000} \\
X^{1^{(k)}}F_0^\sigma+ X^{0^{(k)}}F_1^\sigma &=& \sum_{\tau=1}^{N} Q_{\sigma \tau} F_1^\tau.\label{300}
\end{eqnarray}
\end{theorem}

\begin{remark}\label{rem:triv:BGI}
It is clear from equations \eqref{3000}, \eqref{300} that if $X^0$ is an exact symmetry of equation \eqref{unperturbed eq}, then $X=\epsilon X^0$ is an approximate symmetry of equation \eqref{eq:perturbed:general}. In other words, any symmetry generator $X^0$ of an unperturbed system \eqref{unperturbed eq} yields an approximate symmetry
\beq\label{eq:BGIsymm:trivial}
X=\epsilon X^0
\eeq
of the perturbed system \eqref{eq:perturbed:general}. We call such approximate generators with no $O(\epsilon)$ part \emph{trivial approximate symmetries} of the perturbed system \eqref{eq:perturbed:general}.
\end{remark}

\begin{remark} 
Solving the determining equations \eqref{detapprox} to calculate first-order BGI approximate point symmetry components for equations \eqref{eq:perturbed:general} with a small parameter can be subdivided in the following steps:
\begin{enumerate}
\item Compute an exact point/local symmetry generator $X^0$ of the unperturbed equations \eqref{unperturbed eq} using determining equations \eqref{det:eq:exact} for exact local or point symmetries.
\item Find the corresponding first-order deformation (the part $X^1$ of the generator \eqref{X:X0:X1}) using the equation
\[\label{11}
     X^{1^{(k)}}F_0^\sigma\bigg|_{F_0^\sigma=0}=G(x,v,\partial v,\ldots, \partial^k v),
\]
where $G$ is obtained from the coefficients of $\epsilon$ in
\[
  -X^{0^{(k)}}(F_0^\sigma+\epsilon F_1^\sigma)\bigg|_{F_0^\sigma+\epsilon F_1^\sigma=0}, \quad \sigma=1,...,N.
\]
\end{enumerate}
\end{remark}


\subsection*{\textbf{(D)} Stable and unstable symmetries in the BGI framework}

The determining equations \eqref{detapprox} for the components of the approximate BGI point (or local) symmetry generator \eqref{X:X0:X1} (or  \eqref{X:X0:X1:higher}) may also contain restrictions on the components of the exact local symmetry generator $X^0$. It follows that some exact local symmetries of the unperturbed equations \eqref{unperturbed eq} may disappear from the classification of the approximate BGI symmetries of the perturbed equations \eqref{eq:perturbed:general}. In \cite{ibragimov1995crc}, an exact point (or local) symmetry $X^0$ of the unperturbed equations \eqref{unperturbed eq}  is called \textit{\textrm{stable}} if there exists a point (or local) generator $X^{1}$ such that
\eqref{X:X0:X1} (or respectively \eqref{X:X0:X1:higher}) is an approximate BGI symmetry of the perturbed equation \eqref{eq:perturbed:general}.  If all symmetries of the equations \eqref{unperturbed eq}  are \emph{stable}, the perturbed equations \eqref{eq:perturbed:general} are said to \emph{inherit} the symmetries of the unperturbed equations.

\begin{example}\label{Ex A perturbed}
Consider the second-order ODE
\begin{equation}\label{ypp1}
y''=\epsilon (y')^{-1}
\end{equation}
which is a perturbed version of the ODE \eqref{ypp:0}. The latter has eight exact point symmetries given by \eqref{symm:Ypp0:generators}. Let
\[
      X  =  \left(\xi^0(x,y)+\epsilon \xi^1(x,y)\right)\dfrac{\partial}{\partial x}+  \left(\eta^0(x,y)+\epsilon \eta^1(x,y)\right) \dfrac{\partial}{\partial y}
\]
be the approximate symmetry generator of \eqref{Ex A perturbed}, where $X^0$ is an exact symmetry generator of the unperturbed ODE. The determining equations \eqref{detapprox} for approximate symmetries yield
\begin{equation}\label{eqn10}
\eta^1_{xx}+(2\eta^1_{xy}-\xi^1_{xx})y'+ (\eta^1_{yy}-2\xi^1_{xy}){{y'}^{2}}-\xi^1_{yy}{{y'}^{3}}= (3\xi^0_x-2\eta^0_y)y'^{-1}+4\xi^0_y-\eta^0_x y'^{-2},
\end{equation}
where $\xi^0$, $\eta^0$ are exact symmetry components \eqref{symm:Ypp0} computed in Example \ref{Example A-unperturbed}.
The determining equations \eqref{eqn10} splits into the PDEs
\begin{equation}\label{eq:ExAperturbed:approx:comp}
 \eta^1_{xx}=4C_6,\quad 2\eta^1_{xy}-\xi^1_{xx}=0,\quad \eta^1_{yy}-2\xi^1_{xy}=0,\quad \xi^1_{yy}=0,
\end{equation}
for $\xi^1$, $\eta^1$, and the additional conditions
\[
 3\xi^0_x-2\eta^0_y=0,\quad \eta^0_x=0
\]
on the unperturbed symmetry components $\xi^0$, $\eta^0$ \eqref{symm:Ypp0}. These provide restrictions on free constants in \eqref{symm:Ypp0}:
\[
C_1=C_2=C_3=0,\quad  C_4=\dfrac{3}{2}\,C_7.
\]
The remaining space of exact symmetry components $\xi^0,\,\eta^0$ reduces to
\[
\xi^0 = \frac{2C_4}{3}x+C_6y+C_8,\quad \eta^0 = C_4y+C_5.
\]
The approximate components are found from \eqref{eq:ExAperturbed:approx:comp} and have the form
\beq\label{symm:Ypp0:BGI:point:xi:eta}
\barr
  \xi^1(x,y) &= a_1x^2+\dfrac{a_2}{2}xy+a_3x+a_4y+a_5,\\
  \eta^1(x,y) &=  2C_6x^2+a_1 xy+\dfrac{a_2}{2}y^2+a_6x+a_7y+a_8.
\earr
\eeq
Since the constants $a_1\ldots a_8$ and $C_4, C_5, C_6, C_8$ are free, the ODE \eqref{Ex A perturbed} admits 12 approximate point symmetries. These approximate symmetries can be divided into the following classes:
\begin{subequations}\label{symm:Ypp0:BGI:point:generators:approx:All}
\begin{enumerate}
  \item Exact symmetries inherited from the unperturbed ODE \eqref{ypp:0}, involving only $O(\epsilon^0)$ components
\beq\label{symm:Ypp0:BGI:point:generators:approx1}
 X_{9}= X^0_4+\frac{2}{3}X^0_7,\quad X_{10}= X^0_5,\quad \quad X_{12}= X^0_8.
\eeq

  \item A genuine approximate symmetry
\beq\label{symm:Ypp0:BGI:point:generators:approx2}
X_{11}= X^0_6+2\epsilon x^2 \dfrac{\partial}{\partial y}
\eeq
with $O(\epsilon^0)$ part inherited from the stable symmetry $X^0_6$ of the unperturbed ODE \eqref{ypp:0} (see \eqref{symm:Ypp0:generators}).

  \item Eight trivial symmetries $X_j=\epsilon X^0_j$, $j=1,2,...,8$, given by
\beq\label{symm:Ypp0:BGI:point:generators:triv}
\barr
  X_1 = \epsilon\left(xy\dfrac{\partial}{\partial y}+ x^2\dfrac{\partial}{\partial x}\right),\quad X_2=\epsilon\left(x\dfrac{\partial}{\partial y}\right),\quad X_3= \epsilon\left(\dfrac{y^2}{2}\dfrac{\partial}{\partial
  y}+\dfrac{xy}{2}\dfrac{\partial}{\partial x}\right),  \\[2ex]
  X_4 = \epsilon y\dfrac{\partial}{\partial y},\quad X_5=\epsilon \dfrac{\partial}{\partial y},\quad X_6= \epsilon y\dfrac{\partial}{\partial x},\quad X_7= \epsilon x\dfrac{\partial}{\partial x},\quad
  X_8= \epsilon \dfrac{\partial}{\partial x},\\[2ex]
\earr
\eeq
  corresponding to the free constants $a_1\ldots a_8$ in \eqref{symm:Ypp0:BGI:point:xi:eta}, having only $O(\epsilon)$ components, and arising from each exact point symmetry \eqref{symm:Ypp0:generators} of the unperturbed ODE \eqref{ypp:0}.
\end{enumerate}
\end{subequations}

\medskip Concerning the ``fate" of the exact point symmetries \eqref{symm:Ypp0:generators} of the unperturbed ODE \eqref{ypp:0} in the approximate symmetry classification \eqref{symm:Ypp0:BGI:point:generators:approx:All} of the perturbed ODE \eqref{Ex A perturbed}, it turns out that only four exact symmetries are stable: these are $X^0_5$, $X^0_6$, $X^0_8$ and the linear combination
\[
X^0_s = X^0_4+\frac{2}{3}X^0_7
\]
that is contained in $X_{9}$ of \eqref{symm:Ypp0:BGI:point:generators:approx1}. The other four symmetries of the unperturbed ODE \eqref{ypp:0} are \textit{\textrm{unstable}}, including the generators $X^0_1,\,X^0_2,\,X^0_3$ in \eqref{symm:Ypp0:generators}, and the transverse linear combination of $X^0_4$ and $X^0_7$:
\begin{equation}\label{z}
X^0_u= X^0_4-\frac{3}{2}X^0_7.
\end{equation}
\end{example}

\section{Exact and approximate point symmetries of algebraic equations} \label{sec:Alg}

First we analyze the relationship between exact and approximate point symmetries of algebraic equations. Let $x=(x^1,...,x^n) \in \mathbb{R}^n$, $n\geq 2$.  Let $F_0(x)$ be a sufficiently smooth scalar function. An algebraic equation
\begin{equation}\label{eq2}
  F_0(x)=\const
\end{equation}
defines a family of surfaces (curves) in $\mathbb{R}^n$. A family of perturbed surfaces (curves) is given by
\begin{equation}\label{m11}
  F(x)=F_0(x)+\epsilon F_1(x)=\const.
\end{equation}

\subsection{Exact symmetries of unperturbed and perturbed algebraic equations}\label{sec:plate:localCLs:pot}
The exact symmetry generator of the unperturbed equation $F_0=\const$ is given by
\begin{equation}\label{m12}
  X^0=\sum_{i=1}^{n}\xi^{0i}(x)\dfrac{\partial}{\partial x^i}.
\end{equation}
To find the infinitesimals, we apply the determining equations specifying the condition that every solution curve of \eqref{eq2} is mapped into a solution curve of \eqref{eq2}:
\begin{equation}\label{oo}
   X^0F_0(x)=\sum_{i=1}^{n}\xi^{0i}(x)\dfrac{\partial F_0}{\partial x^i}\equiv 0.
\end{equation}
Assuming without loss of generality that ${\partial F_0}\big/{\partial x^1}\neq 0$, one can solve for
\begin{equation}\label{eq10}
\xi^{01}={-\sum_{i=2}^{n}\xi^{0i}(x)\dfrac{\partial F_0}{\partial x^i}}\bigg/{\dfrac{\partial F_0}{\partial x^1}},
\end{equation}
keeping $\xi^{02}(x),...,\xi^{0n}(x)$ arbitrary functions that parameterise an infinite-parameter Lie algebra of point symmetries of the family of surfaces \eqref{eq2}.  In the same fashion, an exact symmetry generator of the family of perturbed equations \eqref{m11} is given by
\begin{equation}\label{m1444}
  Y=\sum_{i=1}^{n}\eta^{i}(x;\epsilon)\dfrac{\partial}{\partial x^i}.
\end{equation}
Applying the determining equations to find exact symmetries of the perturbed equations \eqref{m11}, one has
\begin{equation}\label{m103}
  YF(x) \sum_{i=1}^{n}\eta^{i}(x;\epsilon)\bigg(\dfrac{\partial F_0}{\partial x^i}+\epsilon \dfrac{\partial F_1}{\partial x^i}\bigg) \equiv 0.
\end{equation}
If at least one of the functions ${\partial F_0}\big/{\partial x^1}$,\, ${\partial F_1}\big/{\partial x^1}$ is nonzero, one can write
\begin{equation}\label{eq12}
  \eta^1(x;\epsilon)={-\sum_{i=2}^{n}\eta^{i}\left(\dfrac{\partial F_0}{\partial x^i}+\epsilon \dfrac{\partial F_1}{\partial x^i}\right) }\bigg/\left({\dfrac{\partial
  F_0}{\partial x^1}}+\epsilon\dfrac{\partial F_1}{\partial x^1}\right)
\end{equation}
in terms of arbitrary functions $\eta^2(x;\epsilon)$, $\ldots$, $\eta^n(x;\epsilon)$ that define the infinite-parameter symmetry generator \eqref{m1444}. From the comparison of \eqref{eq10} and \eqref{eq12}, the following simple theorem is established.
\begin{theorem}
Suppose that the unperturbed algebraic equation \eqref{eq2} admits a point symmetry with infinitesimal generator \eqref{m12}. Then there exists a point symmetry generator \eqref{m1444} of the perturbed equation \eqref{m11} such that $Y\equiv X^0$ when $\epsilon=0$.
\end{theorem}
Indeed, one can take $\eta^{i}(x;\epsilon)= \xi^{0i}$, $i=1,...,n$; then $\eta^1(x;\epsilon)$ \eqref{eq12} matches $ \xi^{01}$ \eqref{eq10} when $\epsilon=0$. It follows that all exact symmetries of the unperturbed equation \eqref{eq2} carry over to the perturbed family \eqref{m11}.

\subsection{BGI approximate symmetries of the perturbed algebraic equation (\ref{m11})}\label{sec:Pr:sym}
Let
\beq \label{alg eqns approx symm}
      X= X^0+\epsilon X^1 = \sum_{i=1}^{n}\xi^{0i}(x)\dfrac{\partial}{\partial x^i} +\epsilon \sum_{i=1}^{n}\xi^{1i}(x)\dfrac{\partial}{\partial x^i}
\eeq
be an BGI approximate point symmetry generator admitted by the family of perturbed surfaces \eqref{m11}, where $X^0$ is the exact symmetry generator of the unperturbed equations \eqref{eq2}. Applying the determining equation
\[
  \left( X^0+\epsilon X^1\right)\left(F_0+\epsilon F_1\right)=o(\epsilon),
\]
we find that the infinitesimals $\xi^{1i}$ satisfy
\begin{equation}\label{m104}
\sum_{i=1}^{n}\xi^{1i}(x)\dfrac{\partial F_0}{\partial x^i}=- \sum_{i=1}^{n}\xi^{0i}(x)\dfrac{\partial F_1}{\partial x^i}.
\end{equation}
As in equation (\ref{eq10}), if ${\partial
F_0}\big/{\partial x_1}\neq 0$, one can solve for
\begin{equation}\label{eq11}
\xi^{11}= -\left({\sum_{i=2}^{n}\xi^{1i}(x)\frac{\partial F_0}{\partial x^i}}{+\sum_{i=1}^{n}\xi^{0i}(x)\frac{\partial F_1}{\partial x^i}}\right)\bigg/{\frac{\partial
F_0}{\partial x^1}},
\end{equation}
where the infinitesimals $\xi^{12}(x)$, $\ldots$, $\xi^{1n}(x)$ are arbitrary functions. The family of perturbed equations \eqref{m11} consequently admits an infinite-parameter approximate symmetry generator
\begin{eqnarray}\label{eq19}
\nonumber 
X &=& \xi^{01}\dfrac{\partial}{\partial x^1}+ \sum_{i=2}^{n}\xi^{0i}(x)\dfrac{\partial}{\partial x^i}+\epsilon \left(\xi^{11}\dfrac{\partial}{\partial x^1}+
\sum_{i=2}^{n}\xi^{1i}(x)\dfrac{\partial}{\partial x^i}\right) \\
&=&  \left(\dfrac{{\epsilon \xi^{01} \dfrac{\partial F_1}{\partial x^1}+\sum \limits_{i=2}^{n}\xi^{0i}\left(\dfrac{\partial F_0}{\partial x^i}+\epsilon \dfrac{\partial
F_1}{\partial x^i} \right)}{+\epsilon \xi^{1i}\dfrac{\partial F_0}{\partial x^i}}} {{{-\partial F_0}\big/{\partial x^1}}}\right)\dfrac{\partial}{\partial
x^1}+\sum_{i=2}^{n}\left(\xi^{0i}+\epsilon \xi^{1i}\right)\dfrac{\partial}{\partial x^i}.
\end{eqnarray}
The following theorem holds.
\begin{theorem}\label{stability of alg eqn}
For each exact symmetry generator \eqref{m12} of the unperturbed algebraic equations \eqref{eq2}, there is a corresponding first-order deformation $X^1$ such that
\eqref{alg eqns approx symm} is an approximate BGI symmetry generator of the family of perturbed equations \eqref{m11}.
\end{theorem}
It follows that every exact point symmetry of the unperturbed algebraic equation \eqref{eq2} is stable, that is, its generator $X_0$ is the $O(\epsilon)$ part of some approximate symmetry generator \eqref{alg eqns approx symm} of the perturbed equation \eqref{m11}. Moreover, due to the presence of additional arbitrary functions
$\xi^{1i}$, $i=2,\ldots,n$, the approximate symmetry generator \eqref{eq19} of the family of perturbed equations \eqref{m11} is more general than the exact symmetry generator \eqref{m1444} of the same. We now consider a simple example in detail.

\begin{figure}[H]
\centering
\begin{subfigure}{0.4\textwidth}\label{fig:ex3-a}
	\includegraphics[width=\textwidth]{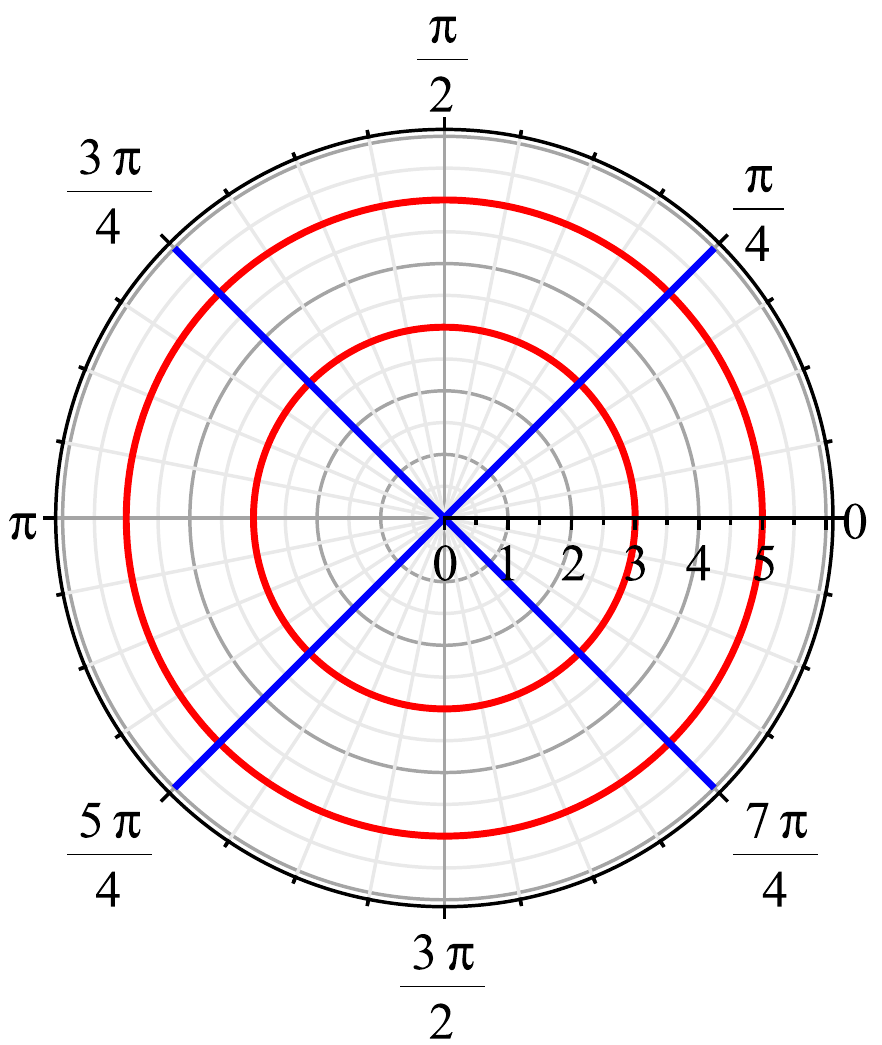}
	\caption{}
\end{subfigure}\hfill
\begin{subfigure}{.4\textwidth}\label{fig:ex3-d}
	\includegraphics[width=\textwidth]{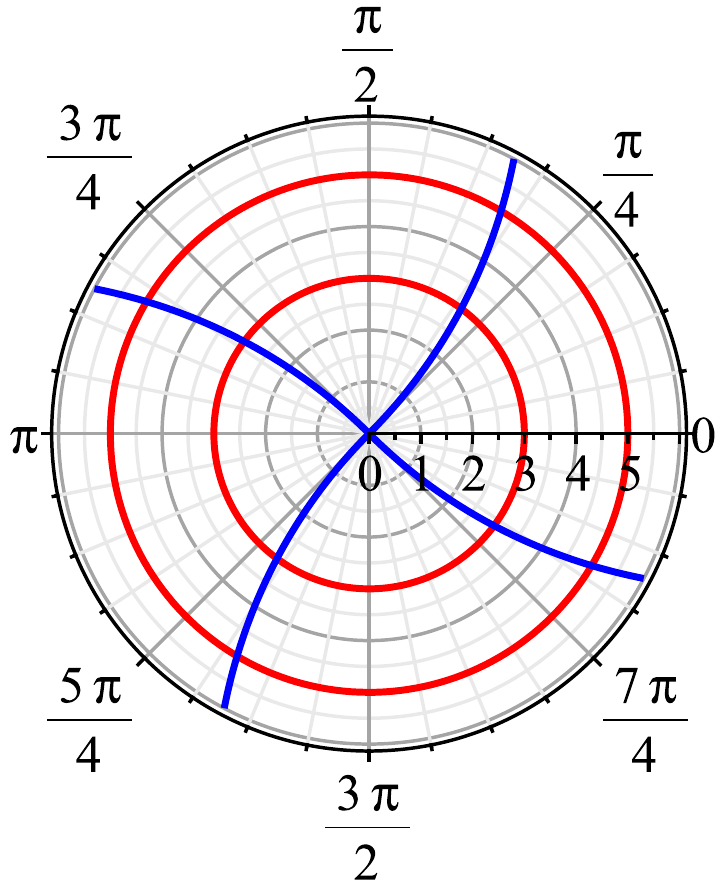}
	\caption{}
\end{subfigure}
\caption[]{
The family of circles \eqref{m1} (a) and their shape under the transformations \eqref{m6} for $a=0.03$ (b). The radial lines are shown in blue for reference.} %
\label{figure1}
\end{figure}

\begin{example}
Consider a family of circles in polar coordinates
\begin{equation}\label{m1}
F_0(r,\theta)=r=\const ,
\end{equation}
and a family of perturbed circles
\begin{equation}\label{m2}
F(r,\theta)= r+ \epsilon e^{-k \theta}=C=\const,
\end{equation}
where $k>0$ is a fixed constant. Let
\end{example}
\[
   X^0=\xi^0(r,\theta)\dfrac{\partial}{\partial r}+ \eta^0(r,\theta) \dfrac{\partial}{\partial \theta}
\]
be the symmetry generator of the family of equations \eqref{m1}. Using the determining equations \eqref{oo}, one gets $\xi^0\equiv 0$, $\eta^0=\eta^0(r,\theta)$. Consequently, all symmetries of the family of circles \eqref{m1} are given by
\begin{equation}\label{m3}
X^0= \eta^0(r,\theta) \dfrac{\partial}{\partial \theta}.
\end{equation}
For example, if $\eta^0=r$, the corresponding global transformation is the one-parameter ($a$) Lie group
\beq\label{m6}
  {r^*} = r, \quad  {\theta^*} = \theta+ar.
\eeq
The equations \eqref{m6} transforms circles to circles and lines to spirals as shown in Figure \ref{figure1}.
For the perturbed circles \eqref{m2}, the exact symmetry generator can be sought in the form
\begin{equation}\label{eq20}
  Y= \eta^1(r,\theta) \dfrac{\partial}{\partial r}+ \eta^2(r,\theta) \dfrac{\partial}{\partial \theta}.
\end{equation}
Using the formula \eqref{eq12}, one gets $\eta^1= \epsilon k e^{-k \theta}\eta^2,$ where $\eta^2(r,\theta)$ is an arbitrary function. Take, for example, $\eta^2=\eta^0=r.$ Then the perturbed equation \eqref{m2} admits the Lie group of transformations:
\beq\label{m7}
  {r^*} =r^*(r,\theta;a,\epsilon)= r+a\epsilon kr e^{-k \theta}+o(a), \quad
  {\theta^*} =\theta^*(r,\theta;a,\epsilon)= \theta+ar+o(a),
\eeq
where
\beq
\dfrac{d{r^*}}{da} = \epsilon k r^* e^{-k {\theta^*}},\quad \dfrac{d{\theta^*}}{da} = r^*,\quad {r^*}\bigg|_{a=0}=r,\quad {\theta^*}\bigg|_{a=0}=\theta.
\eeq
   The above system is equivalent to
   \begin{equation}\label{m8}
     \dfrac{d^2{\theta^*}}{da^2}= \epsilon k  \dfrac{d{\theta^*}}{da} e^{-k {\theta^*}},
   \end{equation}
    which has a solution
   \begin{equation}\label{eq23}
     \theta^*=\dfrac{1}{k} \ln \bigg[\dfrac{kre^{k \theta +a[kr+\epsilon e^{-k {\theta}} ]}+\epsilon}{kr+\epsilon e^{-k {\theta}}}\bigg]
   \end{equation}
   and $r^*$ has the form
   \begin{equation}\label{eq24}
     r^*=\dfrac{[kr+\epsilon e^{-k {\theta}}]re^{k \theta+a[kr+\epsilon e^{-k {\theta}}]}}{kre^{k \theta+a[kr+\epsilon e^{-k {\theta}}]}+\epsilon}.
   \end{equation}
   \begin{figure}[H]
	\begin{subfigure}{.4\textwidth}
		\includegraphics[width=\textwidth]{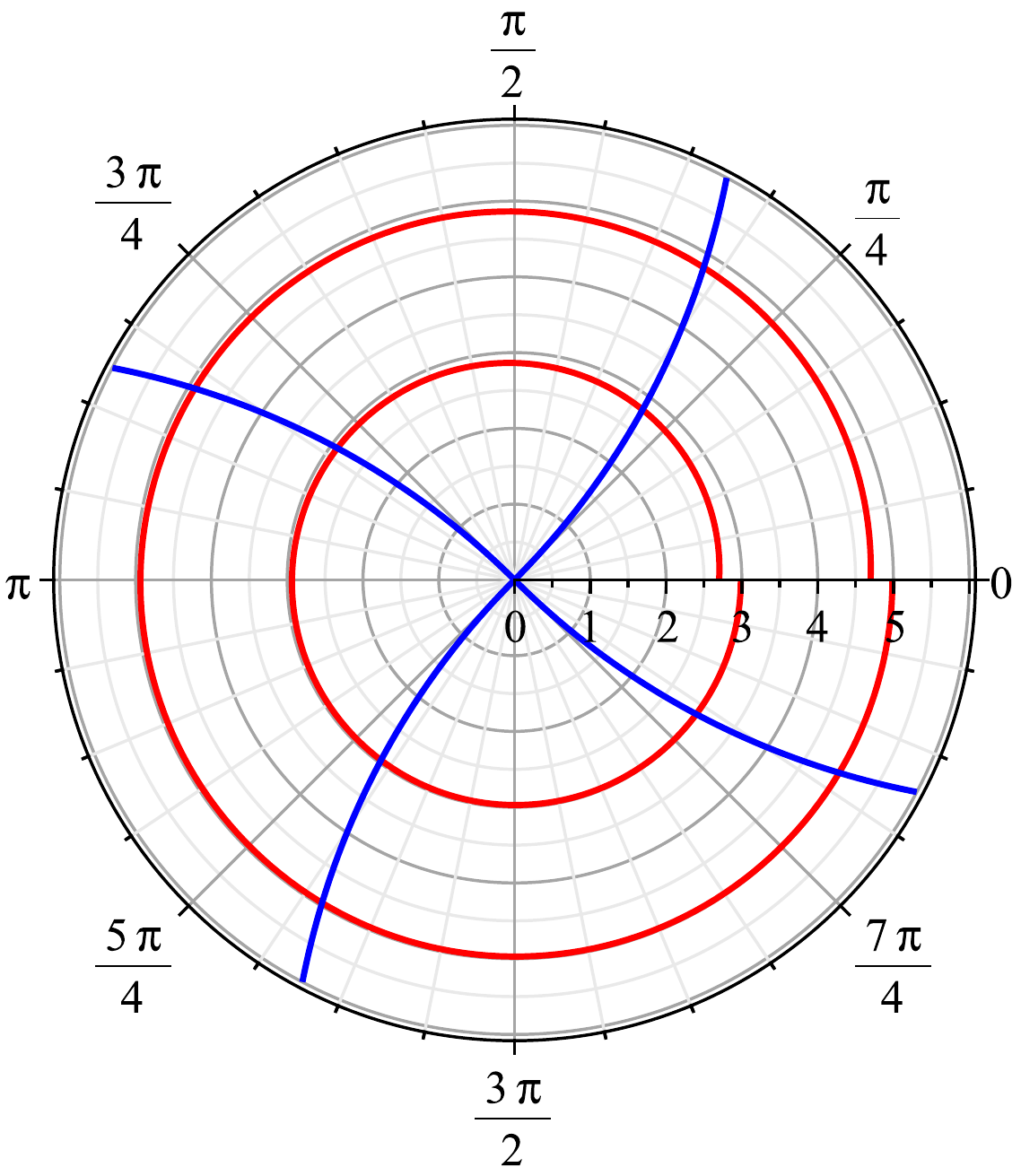}
		\caption{ }\label{fig2a}
	\end{subfigure} \hfill
	\begin{subfigure}{.4\textwidth}
		\includegraphics[width=\textwidth]{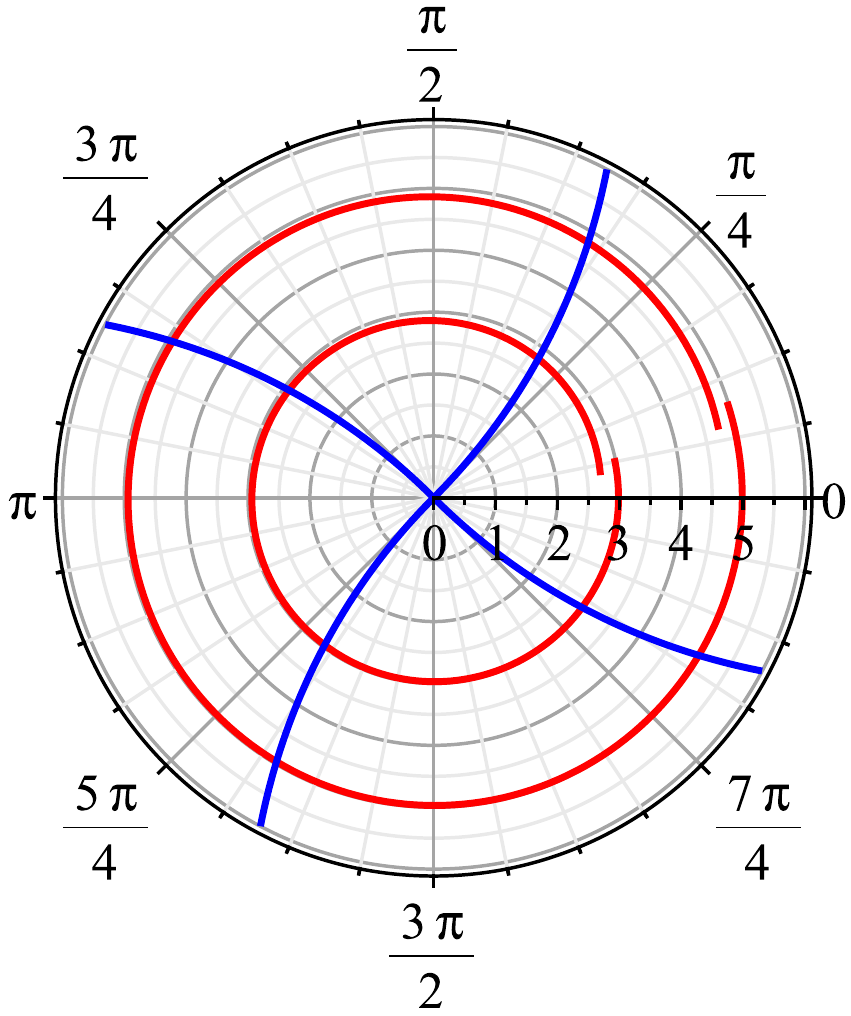}
		\caption{}\label{fig2b}
	\end{subfigure}\\
\centering
\begin{subfigure}{.4\textwidth}
		\includegraphics[width=\textwidth]{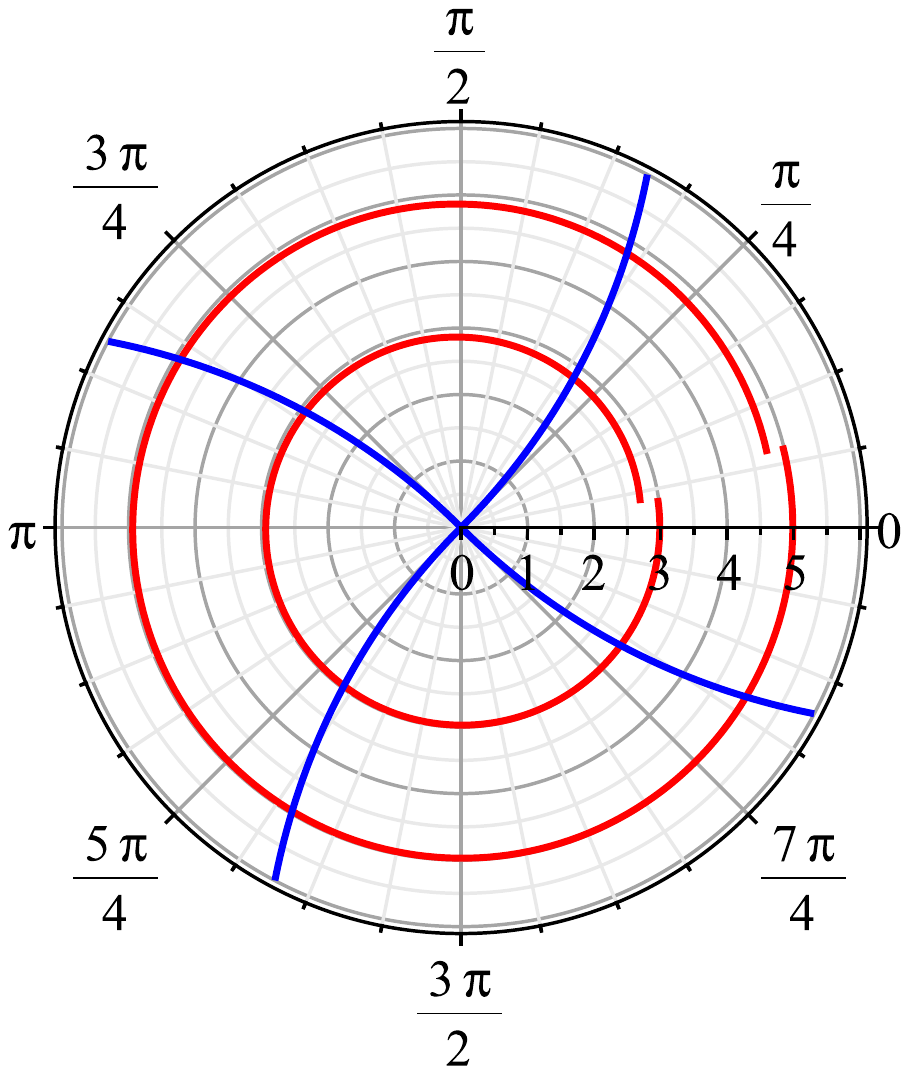}
		\caption{ }\label{fig2c}
	\end{subfigure}
\caption[]{
Family of perturbed circles \eqref{m2} (a) and their graphs under the transformations \eqref{m7}(b) and under the transformations \eqref{eq23}, \eqref{eq24} for $a=0.03$, $k=0.5$, $\epsilon=0.3$ (c).}%
\end{figure}
Figure \ref{fig2a} shows the perturbation to the family of circles \eqref{m1} caused by a small parameter ($\epsilon$). Under the transformations \eqref{m7}, the perturbed circles \eqref{m2} are rotated counter-clockwise as shown in Figure \ref{fig2b}. The action of the exact transformations \eqref{eq23} and \eqref{eq24} on the perturbed circles \eqref{m2} is shown in Figure \ref{fig2c}.

The approximate symmetry generator for the perturbed equation \eqref{m2} has the form
 \begin{equation}\label{algapprox}
   X = X^0+\epsilon X^1 = \left(\xi^0(r,\theta)+\epsilon \xi^1(r,\theta)\right)\dfrac{\partial}{\partial r}+ \left(\eta^0(r,\theta)+\epsilon \eta^1(r,\theta)\right) \dfrac{\partial}{\partial \theta},
 \end{equation}
    where $X^0$ is the exact symmetry generator given by equation \eqref{m3}. From the equation \eqref{eq11}, one finds $\xi^1=ke^{-k \theta}\eta^0 $, $\eta^1=\eta^1(r,\theta).$ Thus, the approximate symmetry generator \eqref{algapprox} becomes
    \begin{equation}\label{m4}
      X = \eta^0\dfrac{\partial}{\partial \theta}+\epsilon \left( ke^{-k \theta}\eta^0\dfrac{\partial}{\partial r} + \eta^1(r,\theta) \dfrac{\partial}{\partial \theta}\right).
    \end{equation}
     The term $\eta^0 \partial/\partial \theta$ in \eqref{m4} corresponds to an exact symmetry of the unperturbed equation \eqref{m1}. It follows that the exact symmetry generator \eqref{m3} of \eqref{m1} is \textit{stable}. By taking $\eta^0=r$, the perturbed circles \eqref{m2} admit Lie group of approximate transformations given by
    \begin{equation}\label{m10}
      r^* = r+a\epsilon kre^{-k \theta}+ o(a),\quad \theta^* = \theta+ar+a \epsilon \eta^1+ o(a),
    \end{equation}
    which coincides with \eqref{m7} when $\eta^1=0$.

\begin{figure}[H]
       \centering
       \includegraphics[width=.4\textwidth]{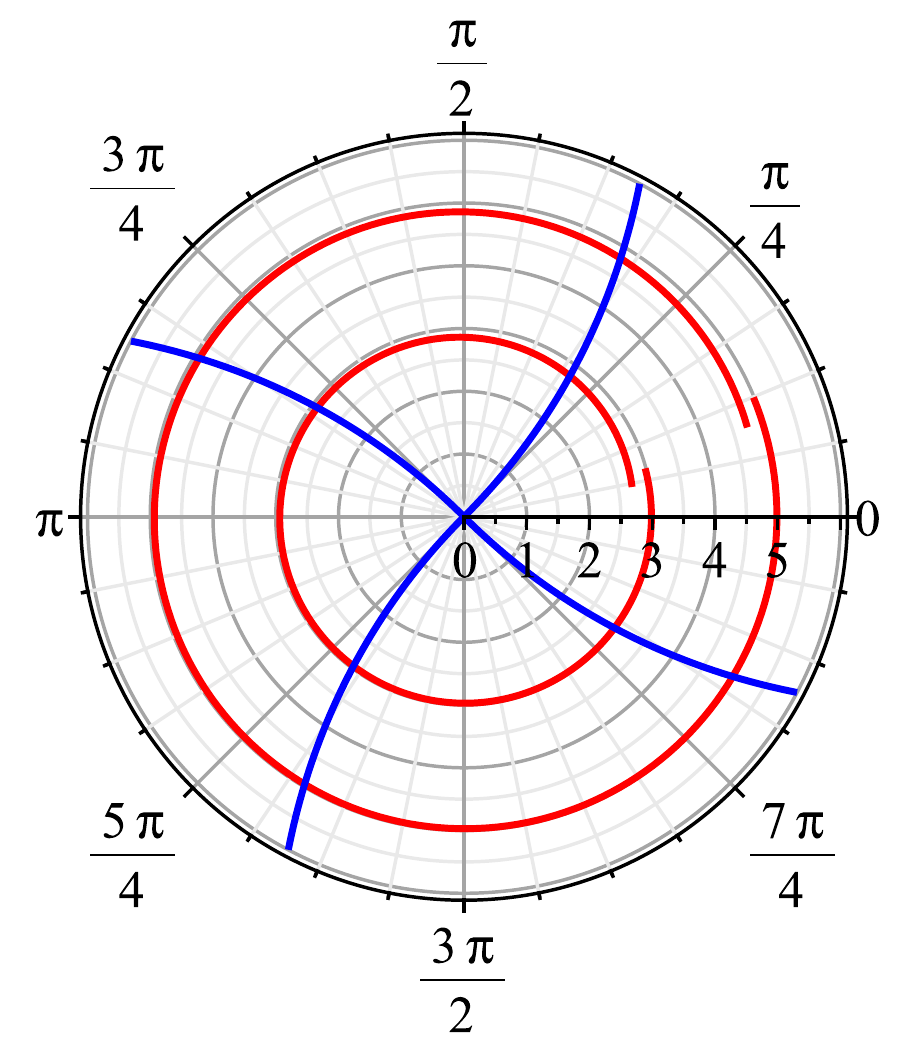}
       \caption{Perturbed circles under the transformation \eqref{m10} for $a=0.03$, $k=0.5$,\,$\epsilon=0.3$ and $\eta^1=r$.}\label{fig3}
     \end{figure}

    In Figure \ref{fig3}, action of \eqref{m10} on the perturbed circles \eqref{m2} is shown when $\eta^1=r\neq 0$. If $\eta^1=0$, Figure \ref{fig3} would coincide with Figure \ref{fig2b}.

\medskip

Summarizing the above results, the following statement can be established.
\begin{proposition}
For any point symmetry $X^0$ \eqref{m12} of an algebraic equation \eqref{eq2}, there exists a corresponding point symmetry $Y$ \eqref{m1444} of the perturbed equation \eqref{m11}. Moreover, in the BGI framework, any point symmetry $X^0$ \eqref{m12} of \eqref{eq2} is stable; there always exists an approximate symmetry $X$ \eqref{alg eqns approx symm} of the perturbed equation \eqref{m11} corresponding to $X^0$.
\end{proposition}

\section{Exact and approximate point symmetries of first-order ODEs}\label{sec:1stODEs}

We now analyze and compare the structures of exact point symmetries of perturbed and unperturbed first-order ODEs, and approximate point symmetries of perturbed ODE models. Let
\begin{equation}\label{s1}
  y'=f_0(x,y)
\end{equation}
denote a first-order ODE, and let
\begin{equation}\label{s2}
  y'=f_0(x,y)+\epsilon f_1(x,y)+o(\epsilon)
\end{equation}
be its perturbation.

\subsection{Exact symmetries of an unperturbed first-order ODE }
  Let $X^0$ be an exact symmetry generator admitted by (\ref{s1}):
  \begin{equation}\label{eq25}
     X^0=\xi^0(x,y)\dfrac{\partial}{\partial x}+ \eta^0(x,y) \dfrac{\partial}{\partial y}.
  \end{equation}
  To find exact point symmetries of (\ref{s1}), one prolongs the exact symmetry generator $X^0$ to the first order:
  \begin{equation}
    X^{0^{(1)}}=\xi^0(x,y)\dfrac{\partial}{\partial x}+ \eta^0(x,y) \dfrac{\partial}{\partial y}+ \eta^{0^{(1)}}(x,y,y') \dfrac{\partial}{\partial y'},
  \end{equation}
  where $\eta^{0^{(1)}}(x,y,y')$ is given by
  \begin{equation}
    \eta^{0^{(1)}}=\eta^0_x+(\eta^0_y-\xi^0_x)y'-\xi^0_y{y'}^{2}.
  \end{equation}
  Applying the determining equation (\ref{det:eq:exact}) to find the exact symmetries of (\ref{s1})
  \[
     X^{0^{(1)}}(y'-f_0(x,y))\bigg|_{y'=f_0(x,y)}=0,
  \]
  one obtains the following linear homogeneous first-order PDE:
  \begin{equation}\label{w3}
    \eta^0_x+\eta^0_yf_0-\eta^0 f_{0_y}-\xi^0f_{0_x}-\xi^0_xf_0-\xi^0_yf_0^2=0
  \end{equation}
 for two unknown functions $\xi^0(x,y)$ and $\eta^0(x,y)$. Taking, for example, $\xi^0(x,y)$ as an arbitrary function, one can find $\eta^0(x,y)$
  from the characteristic system
  \begin{equation}\label{b1}
    \dfrac{dx}{1}=\dfrac{dy}{f_0}=-\dfrac{d \eta^0}{\eta^0 f_{0_y}+\xi^0f_{0_x}+\xi^0_xf_0+\xi^0_yf_0^2}.
  \end{equation}
  Note that the solution of the first characteristic equation is the solution of the differential equation (\ref{s1}) itself. It follows that for any $\xi^0$, one can find multiple $\eta^0$ so that (\ref{eq25}) is a symmetry of (\ref{s1}). In particular, for an arbitrary $\xi^0=\xi^0(x,y)$, it is well known that the choice $\eta^0(x,y)=\xi^0(x,y)f_0(x,y)$ yields a point symmetry of \eqref{s1}.
  \begin{example}\label{ex_unpert_1storder}
    Consider a first order ODE (its perturbed version will be used below)
    \begin{equation}\label{s50}
      y'=x.
    \end{equation}
 For example, for $\xi^0(x,y)=y$, the PDE (\ref{w3}) becomes
  \[
     \eta^0_x+x\eta^0_y=y+x^2.
  \]
 First, take $\eta^0=\xi^0f_0=xy.$ Then
 \begin{equation}
     X^0_1=x\dfrac{\partial}{\partial x}+ xy \dfrac{\partial}{\partial y}
  \end{equation}
  is a point symmetry for the ODE \eqref{s50}. More generally, using the method of characteristics, we have
 \begin{equation}\label{b11}
    \dfrac{dx}{1}=\dfrac{dy}{x}=\dfrac{d \eta^0}{y+x^2},
  \end{equation}
  which has a solution
  \begin{equation}\label{eta0}
   \eta^0(x,y)=xy+A \left( y-\dfrac{x^2}{2}\right),
  \end{equation}
  where $A$ is an arbitrary function of its argument. The symmetry generator corresponding to the choice $\xi^0=y$ is given by
  \begin{equation}\label{}
    X^0_2= y\dfrac{\partial}{\partial x}+ \left(xy+A \left( y-\dfrac{x^2}{2}\right)\right) \dfrac{\partial}{\partial y}.
  \end{equation}
  \end{example}
\subsection{Exact symmetries of perturbed first-order ODEs (\ref{s2})}
 Let
 \begin{equation}\label{Y:1st ode}
   Y=\xi(x,y;\epsilon)\dfrac{\partial}{\partial x}+ \eta(x,y;\epsilon) \dfrac{\partial}{\partial y}
 \end{equation}
 be an exact symmetry generator of the perturbed equation (\ref{s2}). The prolongation of the symmetry generator $Y$ is given by
 \[
   Y^{(1)}=Y+\eta^{(1)}(x,y,y';\epsilon)\dfrac{\partial}{\partial y'},
 \]
  where
 \[
  \eta^{(1)}=\eta_x+(\eta_y-\xi_x)y'-\xi_{y}y'{^2}.
 \]
 The determining equation (\ref{det:eq:exact})
 \[
    Y^{{(1)}}(y'-f_0-\epsilon f_1)\bigg|_{y'=f_0+\epsilon f_1}=0
 \]
 yields the following PDE
 \begin{equation}\label{s3}
  \eta_x+(f_{0}+\epsilon f_{1})\eta_y-\eta(f_{0_y}+\epsilon f_{1_y}) -(f_{0_x}+\epsilon f_{1_x})\xi-(f_{0}+\epsilon f_{1})\xi_x-(f_{0}+\epsilon f_{1})^2\xi_y=0.
 \end{equation}
Again, for an arbitrary $\xi(x,y)$, one can obtain the following characteristic system to solve for $\eta(x,y;\epsilon)$:
 \begin{equation}\label{b2}
    \dfrac{dx}{1}=\dfrac{dy}{f_{0}+\epsilon f_{1}}=-\dfrac{d \eta}{\eta(f_{0_y}+\epsilon f_{1_y}) +(f_{0_x}+\epsilon f_{1_x})\xi+(f_{0}+\epsilon f_{1})\xi_x+(f_{0}+\epsilon
    f_{1})^2\xi_y}.
  \end{equation}
  If $\xi=\xi(x,y;\epsilon)$ depends on $\epsilon$ analytically, then so does $\eta=\eta(x,y;\epsilon)$; when $\epsilon=0$, the symmetry \eqref{Y:1st ode} of the perturbed ODE \eqref{s2} reduces to the exact point symmetry \eqref{eq25} of the unperturbed ODE \eqref{s1}. In particular, note that $\eta(x,y;\epsilon)=\xi(x,y;\epsilon)(f_{0}(x,y)+\epsilon f_{1}(x,y))$ solves the PDE \eqref{s3}, where $\xi$ is an arbitrary function. Hence, by taking an arbitrary $\xi$ with $\xi(x,y;0)=\xi^0$, one obtains $\eta$ with $\eta(x,y;0)=\eta^0$.

\begin{example}
Consider a perturbed version of the ODE \eqref{s50}:
\begin{equation}\label{s44}
y'=x+\epsilon y.
\end{equation}
From Example \ref{ex_unpert_1storder},
\begin{equation}\label{X0:ex4.2}
X^0=y\dfrac{\partial}{\partial x}+ xy \dfrac{\partial}{\partial y}
\end{equation}
is an exact symmetry generator for the unperturbed ODE \eqref{s50}. By taking\, $\xi(x,y;\epsilon)=y+\epsilon x$,
one gets $\eta(x,y;\epsilon)=xy+\epsilon (x^2+y^2)+\epsilon^2 xy$. Therefore
\begin{equation}\label{Y:ex4.2}
Y=(y+\epsilon x)\dfrac{\partial}{\partial x}+ (xy+\epsilon (x^2+y^2)+\epsilon^2 xy) \dfrac{\partial}{\partial y}
\end{equation}
is an exact symmetry generator for the perturbed ODE \eqref{s44}. When $\epsilon=0$, the symmetry \eqref{Y:ex4.2} of the perturbed ODE \eqref{s44} reduces to the point symmetry \eqref{X0:ex4.2} of the unperturbed ODE \eqref{s50}.
\end{example}

\subsection{Approximate symmetries of perturbed first-order ODEs (\ref{s2})}
   The approximate symmetry generator admitted by a perturbed ODE (\ref{s2}) has the form
   \begin{equation}\label{X:1st ode}
      X = X^0+\epsilon X^1= \left(\xi^0(x,y)+\epsilon \xi^1(x,y)\right)\dfrac{\partial}{\partial x}+ \left(\eta^0(x,y)+\epsilon \eta^1(x,y)\right) \dfrac{\partial}{\partial y},
   \end{equation}
   and its prolongation is given by
   \begin{equation}\label{}
      X^{(1)} = X^{0^{(1)}}+\epsilon X^{1^{(1)}}= X^{0^{(1)}}+\epsilon(X^1+\eta^{1^{(1)}}(x,y,y') \dfrac{\partial}{\partial y'}),
   \end{equation}
  where $\eta^{1^{(1)}}=\eta^1_x+(\eta^1_y-\xi^1_x)y'-\xi^1_y{y'}^{2}.$ Applying the determining equation for the approximate symmetries of (\ref{s2}), one obtains
  \begin{equation}\label{s5}
     \eta^1_x+(\eta^1_y-\xi^1_x)f_0-\xi^1_y{f_0}^{2}-\xi^1f_{0_x}-\eta^1f_{0_y}=(\xi^0_x-\eta^0_y)f_1+2\xi^0_yf_0f_1+\xi^0f_{1_x}+\eta^0f_{1_y}.
  \end{equation}
  The above equation (\ref{s5}) is a linear nonhomogeneous PDE in two unknowns $\xi^1$ and $\eta^1$. It follows that, all exact symmetries of the unperturbed ODE \eqref{s1} are \textit{\textrm{stable}}. To solve the PDE \eqref{s5}, one can pick an arbitrary function \,$\xi^1$, and solve the resulting PDE for $\eta^1$. In particular, $(\xi^1,\eta^1)$ with $\eta^1=\xi^0f_1+\xi^1f_0$ and $\xi^1$ an arbitrary function are approximate symmetry components corresponding to the exact symmetry generator \eqref{eq25} with $\eta^0=\xi^0f_0$.
  \begin{remark}\label{exact_approx_1st ode_remark}
    If the components $(\xi,\eta)$ of the exact symmetry generator $Y$ \eqref{Y:1st ode} are analytic in $\epsilon$, then the approximate symmetry generator $X$ \eqref{X:1st ode} is contained as the first two terms of the Taylor expansion of $Y$ in $\epsilon$.
  \end{remark}
  \begin{example}
    Consider the perturbed ODE \eqref{s44}. For the exact symmetry generator \eqref{X0:ex4.2} of the unperturbed ODE \eqref{s50}, one can find the approximate symmetry components by first taking an arbitrary value for $\xi^1$, say $\xi^1(x,y)=x$. Then $\eta^1(x,y)$ has the form $\eta^1=\xi^0f_1+\xi^1f_0=x^2+y^2$. Hence
    \begin{equation}\label{approx_ex4.3}
      X=(y+\epsilon x)\dfrac{\partial}{\partial x}+\left (xy+\epsilon (x^2+y^2)\right) \dfrac{\partial}{\partial y}
    \end{equation}
   is an approximate symmetry for \eqref{s44}. Note that by this choice of $\xi^1$, $X$ \eqref{approx_ex4.3} is contained in the exact symmetry generator \eqref{Y:ex4.2}.
  \end{example}
  \subsection{Summary}
  The following statement summarizes the above results.
  \begin{proposition}
    If the components of the exact symmetry generator $Y$ \eqref{Y:1st ode} of the perturbed first-order ODE \eqref{s2} are analytic in $\epsilon$, then the approximate symmetry generator $X$ \eqref{X:1st ode} is contained in the Taylor expansion of $Y$ in terms of $\epsilon$. Furthermore, when $\epsilon=0$, $Y$ reduces to the exact symmetry generator $X^0$ \eqref{eq25} of the unperturbed first-order ODE \eqref{s1}. In the BGI framework, all point symmetries of the unperturbed ODE \eqref{s1} are stable.
  \end{proposition}

\section{Exact and approximate point symmetries of higher-order ODEs}\label{sec:higherODEs}
Here we discuss the BGI approximate symmetries of a perturbed higher-order ODE, and the stability of the exact point symmetries of the unperturbed model. Consider the unperturbed higher-order ODE
  \begin{equation}\label{s06}
    y^{(n)}=f_0(x,y,y',...,y^{(n-1)}),\quad n\geq 2,
  \end{equation}
  and its perturbed version
   \begin{equation}\label{s6}
    y^{(n)}=f_0(x,y,y',...,y^{(n-1)})+\epsilon f_1(x,y,y',...,y^{(n-1)})+o(\epsilon).
  \end{equation}
  \subsection{Exact point symmetries of the unperturbed ODE \eqref{s06}}
  The exact symmetry generator for the unperturbed ODE \eqref{s06} has the form
  \begin{equation}\label{exact genarator}
    X^{0}=\xi^0(x,y)\dfrac{\partial}{\partial x}+ \eta^0(x,y) \dfrac{\partial}{\partial y},
  \end{equation}
  and the $n^{\rm th}$ prolongation of this operator is given by
  \[
    X^{0^{(n)}}=\xi^0(x,y)\dfrac{\partial}{\partial x}+ \eta^0(x,y) \dfrac{\partial}{\partial y}+ \eta^{0^{(1)}}(x,y,y') \dfrac{\partial}{\partial
    y'}+...+\eta^{0^{(n)}}(x,y,y',...,y^{(n)}) \dfrac{\partial}{\partial y^{(n)}},
  \]
  where the extended infinitesimals $\eta^{0^{(k)}}$ satisfy the recursion formula
  \begin{equation}\label{s10}
    \eta^{0^{(k)}}=D\eta^{0^{(k-1)}}-y^{(k)}D\xi^0,\, k\geq1,
  \end{equation}
  where $\eta^{0^{(0)}}=\eta^0$, and $D$ is the total derivative operator \eqref{DO} given by
  \begin{equation}\label{hj}
     D=\dfrac{\partial}{\partial x}+y'\dfrac{\partial}{\partial y}+...+y^{(n+1)}\dfrac{\partial}{\partial y^{(n)}}.
  \end{equation}
  The determining equations for exact symmetries of the unperturbed ODE \eqref{s06}
   \begin{equation}\label{det}
      X^{0^{(n)}}(y^{(n)}-f_0)\bigg|_{y^{(n)}=f_0}=0
   \end{equation}
   yield
   \begin{equation}\label{det1}
     \eta^{0^{(n)}}\bigg|_{y^{(n)}=f_0}-\sum_{k=1}^{n-1}\eta^{0^{(k)}}\frac{\partial f_0}{\partial y^{(k)}}-\xi^0 f_{0_x}-\eta^0 f_{0_y}=0,
   \end{equation}
   which is a linear PDE system on $(\xi^0,\eta^0).$ A sample point symmetry computation for $y''=0$ \eqref{ypp:0} is presented in Example \ref{Example A-unperturbed}.
   \subsection{Exact and approximate point symmetries of the perturbed ODE \eqref{s6}}
  A perturbed ODE \eqref{s6} generally has fewer exact point and local symmetries than the unperturbed ODE \eqref{s06}. Example \ref{ex:exact-of-perturbed} for the ODE $y''=\epsilon (y')^{-1}$ illustrates this trend.

   The approximate symmetry generator of the perturbed ODE (\ref{s6}) is given by
   \beq
    X = X^0+\epsilon X^1=\left(\xi^0(x,y)+\epsilon \xi^1(x,y)\right)\dfrac{\partial}{\partial x}+ \left(\eta^0(x,y)+\epsilon \eta^1(x,y)\right) \dfrac{\partial}{\partial y},
   \eeq
   and its prolongation is given by
   \begin{equation}\label{}
      X^{{(n)}}=\left(\xi^0+\epsilon \xi^1\right)\dfrac{\partial}{\partial x}+ \left(\eta^0+\epsilon \eta^1\right) \dfrac{\partial}{\partial y}+\left(\eta^{0^{(1)}}+ \epsilon \eta^{1^{(1)}}\right)
      \dfrac{\partial}{\partial y'}+...+\left(\eta^{0^{(n)}}+ \epsilon \eta^{1^{(n)}}\right) \dfrac{\partial}{\partial y^{(n)}}.
   \end{equation}
   To find the approximate symmetries of the perturbed ODE (\ref{s6}), we apply the determining equation
   \begin{equation}\label{s7}
     X^{1^{(n)}}(y^{(n)}-f_0)\bigg|_{y^{(n)}=f_0}=-\frac{1}{\epsilon}X^{0^{(n)}}(y^{(n)}-f_0-\epsilon f_1)\bigg|_{y^{(n)}=f_0+\epsilon f_1},
    \end{equation}
equivalent to
\begin{multline}\label{00}
  \left(\eta^{1^{(n)}}-\sum_{k=1}^{n-1} \eta^{1^{(k)}}\frac{\partial f_0}{\partial y^{(k)}}\right)\bigg|_{y^{(n)}=f_0}-\xi^1 f_{0_x}-\eta^1 f_{0_y}=\\
  -\frac{1}{\epsilon}\left(\eta^{0^{(n)}}-\sum_{k=1}^{n-1} \eta^{0^{(k)}}(\frac{\partial f_0}{\partial y^{(k)}}+\epsilon \frac{\partial f_1}{\partial y^{(k)}})-\xi^0
  (f_{0_x}+\epsilon f_{1_x})-\eta^0 (f_{0_y}+\epsilon f_{1_y}) \right)\bigg|_{y^{(n)}=f_0+\epsilon f_1}.
\end{multline}
Note that ${\eta}^{0^{(n)}}$ is linear in $y^{(n)}$, and satisfies the equation
\[
   \eta^{0^{(n)}}= D^n \eta^0 - \sum_{j=0}^{n-1} \binom{n}{j} D^j y' D^{n-j}\xi^0.
\]
Hence the general form for the determining equation for approximate
symmetries of the perturbed ODE  \eqref{s6} is
\begin{multline}\label{000}
    \left(\eta^{1^{(n)}}-\sum_{k=1}^{n-1} \eta^{1^{(k)}}\frac{\partial f_0}{\partial y^{(k)}}\right)\bigg|_{y^{(n)}=f_0}-\xi^1 f_{0_x}-\eta^1
    f_{0_y}=\\(n\xi^0_x-\eta^0_y)f_1+(n+1)y'\xi^0_yf_1+\sum_{k=1}^{n-1} \eta^{0^{(k)}} \frac{\partial f_1}{\partial y^{(k)}}+\xi^0 f_{1_x}+\eta^0 f_{1_y}.
\end{multline}
After replacing \, $y^{(n)}$ by $f_0(x,y,y',..., y^{(n-1)}$), equation \eqref{000} constitutes of differential functions in\, $y',y'',\ldots, y^{(n-1)}$, whose coefficients are the unknown functions \,$\xi^1$,\, $\eta^1$, the unperturbed symmetry components \,$\xi^0$,\, $\eta^0$, and their partial derivatives up to $n^{\rm th}$ order. Hence equation \eqref{000} splits into overdetermined system of PDEs in \,$\xi^1$,\, $\eta^1$, with or without additional conditions on the unperturbed symmetry components \,$\xi^0$,\, $\eta^0$. When such additional conditions are present, an exact symmetry of the unperturbed ODE \eqref{s06} may disappear from the approximate symmetry classification of the perturbed ODE \eqref{s6}, thus becoming unstable (see Example \ref{Ex A perturbed}). The following example illustrates the case when there are no restrictions on the unperturbed symmetry components which leads to the {\textrm{stability}} of all point symmetries of the unperturbed equation.
\begin{example}\label{ex:all:stable}
       Consider a second order ODE
       \begin{equation}\label{eq101}
         y''=\epsilon y'
       \end{equation}
       that is also a perturbed version of \eqref{ypp:0}. Equation \eqref{000} for approximate symmetries  of \eqref{eq101} reads
       \begin{equation}\label{ex2sec5eq1}
         \eta^1_{xx}+(2\eta^1_{xy}-\xi^1_{xx})y'+(\eta^1_{yy}-2\xi^1_{xy})(y')^2-\xi^1_{yy}(y')^3= \eta^0_{x}+\xi^0_{x} y'+2\xi^0_{y} y'^2,
       \end{equation}
       where \,$\xi^0$,\, $\eta^0$ are the unperturbed symmetry components \eqref{symm:Ypp0}. Obviously, equation \eqref{ex2sec5eq1} splits into the following system of PDEs in \,$\xi^1$,\, $\eta^1$
      \begin{equation}\label{w2}
       \eta^1_{xx} = C_1y+C_2, \quad 2\eta^1_{xy}-\xi^1_{xx} = 2C_1x+\dfrac{1}{2} C_3 y+C_7, \quad \eta^1_{yy}-2\xi^1_{xy} = C_3x+2C_6, \quad \xi^1_{yy} = 0,
      \end{equation}
      with no change on \,$\xi^0$,\, $\eta^0$. Solving the above system of PDEs yields the following values of $\xi^1$, and $\eta^1$
    \begin{equation}
      \xi^1=-\dfrac{C_3}{4}x^2y-C_6xy-\dfrac{C_7}{2}x^2+a_1x^2+\dfrac{a_2}{2}xy+a_3x+a_4y+a_5,
    \end{equation}
    \begin{equation}
      \eta^1=\dfrac{C_1}{2}x^2y+\dfrac{C_2}{2}x^2+a_1xy+\dfrac{a_2}{2}y^2+a_6x+a_7y+a_8,
    \end{equation}
    where $a_i$ are arbitrary constants. Consequently, the perturbed ODE \eqref{eq101} admits 16 approximate symmetries given by
    \begin{eqnarray}\label{eqq}
     \nonumber 
      X_1 &=& x^2\dfrac{\partial}{\partial x}+\left(xy+\epsilon \dfrac{x^2y}{2}\right)\dfrac{\partial}{\partial y},\quad X_2=x\dfrac{\partial}{\partial y}+\epsilon
      \dfrac{x^2}{2}\dfrac{\partial}{\partial y},\quad X_3= \left(\dfrac{xy}{2}-\epsilon \dfrac{x^2y}{4}\right)\dfrac{\partial}{\partial
      x}+\dfrac{y^2}{2}\dfrac{\partial}{\partial y},\\
      \nonumber
       X_4 &=& \left(y-\epsilon xy\right) \dfrac{\partial}{\partial x},\quad X_5=\left(x-\epsilon\dfrac{x^2}{2}\right)\dfrac{\partial}{\partial x},\quad X_6=
       y\dfrac{\partial}{\partial y},\quad X_7= \dfrac{\partial}{\partial y},\quad  X_8= \dfrac{\partial}{\partial x}, \\
       \nonumber
      X_9&=& \epsilon \left( xy\dfrac{\partial}{\partial y}+x^2 \dfrac{\partial}{\partial x}\right),\quad X_{10}=\epsilon
      x\dfrac{\partial}{\partial y},\quad X_{11}=\epsilon \left(xy\dfrac{\partial}{\partial x}+y^2\dfrac{\partial}{\partial y}\right),\quad X_{12}= \epsilon y \dfrac{\partial}{\partial y},\\
      X_{13}&=& \epsilon  \dfrac{\partial}{\partial y},\quad X_{14}=\epsilon y\dfrac{\partial}{\partial x},\quad
       X_{15}=\epsilon x\dfrac{\partial}{\partial x},\quad X_{16}=\epsilon \dfrac{\partial}{\partial x}.
    \end{eqnarray}
    All exact symmetries \eqref{symm:Ypp0:generators} of the unperturbed ODE \eqref{ypp:0} are inherited by the approximate symmetries (\ref{eqq}), and thus are {\textrm{stable}} by definition. Note that the symmetries $X_9$, $X_{10}$,\ldots, $X_{16}$ are trivial symmetries arising from the point symmetries \eqref{symm:Ypp0:generators} of the unperturbed ODE \eqref{ypp:0}. $X_6$, $X_7$, $X_8$ are exact symmetries directly carrying over from the unperturbed ODE \eqref{ypp:0}, while $X_1$, $X_2$, ..., $X_5$ are genuine approximate symmetries with $O(\epsilon^0)$ parts inherited from the exact point symmetries of the unperturbed ODE \eqref{ypp:0}.
     \end{example}

\section{Exact and approximate local symmetries of higher-order ODEs}\label{sec:localsym}
     It is natural to expect that under a perturbation of an ODE model, an exact local symmetry of the original system would become an approximate local symmetry of a perturbed ODE system. This, however, is not the case for point symmetries, as some of those are {\textrm{unstable}}, and totally disappear from the classification of approximate symmetries of the perturbed model.  In the current section, we show that a point symmetry of unperturbed ODEs corresponds to higher-order approximate symmetry of a perturbed model.
\subsection{Exact local symmetries of the unperturbed ODE \eqref{s06}}
  The local infinitesimal generator \eqref{highgenerator1} for an unperturbed ODE \eqref{s06} has the form
  \begin{equation}\label{local symm}
    \hat{X}^{0}=\zeta^0(x,y,y',y'',...,y^{(s)})\dfrac{\partial}{\partial y},\quad 1\leq s\leq n-1.
  \end{equation}
  When $s=1$, the local symmetry generator \eqref{local symm} corresponds to the point symmetry generator \eqref{exact genarator} of ODE \eqref{s06} provided that $\zeta^0(x,y,y')=\eta^0(x,y)-y'\xi^0(x,y)$. If $\zeta^0(x,y,y')$ is not linear in $y'$, then \eqref{local symm} corresponds to a contact symmetry generator of ODE \eqref{s06}. When $s\geq2$, the local symmetry generator \eqref{local symm} corresponds to a higher-order symmetry generator of ODE \eqref{s06}.
  The $n^{\rm th}$ prolongation of \eqref{local symm} is given by
  \[
    \hat{X}^{0^{(n)}}=\zeta^0\dfrac{\partial}{\partial y}+\zeta^{0^{(1)}}\dfrac{\partial}{\partial y'}+...+\zeta^{0^{(n)}} \dfrac{\partial}{\partial y^{(n)}},
  \]
  with
  \begin{equation}\label{eq3}
    \zeta^{0^{(j)}}=D^j\zeta^0,\quad j=1,2,...,n,
  \end{equation}
  where $D$ is given by (\ref{hj}). The determining equation for the exact symmetries of the unperturbed equation is
  \begin{equation}\label{z7}
   \hat{X}^{0^{(n)}}(y^{(n)}-f_0)\bigg|_{y^{(n)}=f_0}=0,
   \end{equation}
   or in detail,
   \begin{equation}\label{eq4}
    \zeta^{0^{(n)}}\bigg|_{y^{(n)}=f_0}-\sum_{k=1}^{n-1}\left(\zeta^{0^{(k)}}\frac{\partial f_0}{\partial y^{(k)} }\right) \bigg|_{y^{(n)}=f_0}- \zeta^0 f_{0_y}=0.
   \end{equation}
   The latter is equivalent to
   \begin{equation}\label{eq05}
    D^n \zeta^{0}\bigg|_{y^{(n)}=f_0}=\sum_{k=1}^{n-1}\left(D^{k}\zeta^{0}\frac{\partial f_0}{\partial y^{(k)}}\right) \bigg|_{y^{(n)}=f_0}+ \zeta^0 f_{0_y}.
   \end{equation}
   If $s=n-1$, equation (\ref{eq05}) is a linear homogeneous PDE for $\zeta^0$ with independent variables\, $x,y,y',\ldots, y^{(n-1)}$. This PDE can be written in solved form
   \begin{equation}\label{solved form}
     \dfrac{\partial^n \zeta^0 }{\partial x^n}=R(x,y,y',..., y^{(n-1)},\zeta^0,\partial \zeta^0,...,\partial^n \zeta^0)
   \end{equation}
   for the highest derivative of $\zeta^0$ with respect to the independent variable $x$, where all derivatives with respect to $x$ appearing in the right-hand side of \eqref{solved form} are of lower order than those appearing on the left-hand side. Hence, it is in \textrm{{Cauchy-Kovalevskaya form}} with respect to $x$. It follows that the PDE (\ref{eq05}) is solvable when $s=n-1$. When $s< n-1$, equation (\ref{eq05}) splits into an overdetermined system of linear homogeneous PDEs which has at most a finite number of linearly independent solutions (e.g.,\cite{bluman2008symmetry}).
   \subsection{Approximate local symmetries of the perturbed ODE \eqref{s6}}
   The higher-order approximate symmetry generator for the ODE \eqref{s6} is given by
   \begin{equation}\label{local_approx}
      \hat{X} =\hat{ X}^0+\epsilon \hat{X}^1=\left(\zeta^0(x,y,y',...,y^{(s)})+\epsilon \zeta^1(x,y,y',...,y^{(\ell)})\right)\dfrac{\partial}{\partial y}.\quad s,\ell \leq n-1.
   \end{equation}
   The prolongation of this generator has the form
   \begin{equation}\label{}
     \hat{X}^{(n)} = \hat{X}^{0^{(n)}}+\epsilon \hat{X}^{1^{(n)}}= \hat{X}^{0^{(n)}}+\epsilon \big(X^1+\zeta^{1^{(1)}}\dfrac{\partial}{\partial y'}+...+\zeta^{1^{(n)}} \dfrac{\partial}{\partial y^{(n)}}\big),
   \end{equation}
   with  $\zeta^{1^{(j)}}=D^j\zeta^1,\, j=1,2,...,n.$ To find the approximate symmetries of the perturbed ODE \eqref{s6}, we apply the determining equations for approximate symmetries
   \begin{equation}\label{det2}
      \hat{X}^{(n)}(y^{(n)}-f_0- \epsilon f_1)\bigg|_{y^{(n)}=f_0 + \epsilon f_1}=o(\epsilon).
   \end{equation}
   First, one computes an exact local symmetry generator \eqref{local symm} of the unperturbed ODE \eqref{s06}. Then, the first-order deformation $\hat{X}^1$ can be found from the equation
   \[
     \hat{X}^{1^{(n)}}(y^{(n)}-f_0)\bigg|_{y^{(n)}=f_0}=G(x,y,y',...,y^{(n-1)}),
\]
   where $G$ is the coefficient of $\epsilon$ in
   \begin{equation}\label{eq009}
     -\left(\hat{X}^{0^{(n)}}(y^{(n)}-f_0- \epsilon f_1)\right)\bigg|_{y^{(n)}=f_0+\epsilon f_1}.
   \end{equation}
 The determining equation \eqref{det2} becomes
   \begin{equation}\label{eq005}
    D^n \zeta^{1}\bigg|_{y^{(n)}=f_0}-\sum_{k=1}^{n-1}\left(D^{k}\zeta^{1}\frac{\partial f_0}{\partial y^{(k)}}\right) \bigg|_{y^{(n)}=f_0}- \zeta^1 f_{0_y}=G.
   \end{equation}
   \begin{remark}
     When $\ell=n-1$, equation \eqref{eq005} is a linear nonhomogeneous PDE in $\zeta^1$, and it is in \textrm{{Cauchy-Kovalevskaya form}} with respect to the independent variable $x$, so it has solutions obtainable (at least implicitly) by the method of characteristics. If $\ell=n-1$, any solution of the PDE (\ref{eq005}) has no conditions on the unperturbed symmetry components $\zeta^0$. The following theorem holds.
   \end{remark}
   \begin{theorem}\label{h.o.approx}
     For each exact point or local symmetry \eqref{local symm} of an unperturbed ODE \eqref{s06}, there is an approximate symmetry \eqref{local_approx} of the perturbed ODE \eqref{s6}, with the symmetry component $\zeta^1$ being of order at most $n-1$.
   \end{theorem}
   \subsection{First detailed example}\label{sec:6:3}
  Consider the second-order ODE
  \begin{equation}\label{h101}
    y''=\epsilon {(y')^{-1}}
  \end{equation}
  The ODE \eqref{h101} admits 12 approximate symmetries (see Example \ref{Ex A perturbed} above), where these approximate symmetries do not inherit the following exact symmetries of $y''=0$:
  \[
     X_1^0 = xy\dfrac{\partial}{\partial y}+ x^2\dfrac{\partial}{\partial x},\quad X_2^0=x\dfrac{\partial}{\partial y},\quad X_3^0= \dfrac{y^2}{2}\dfrac{\partial}{\partial
     y}+\dfrac{xy}{2}\dfrac{\partial}{\partial x},\quad X^0_u= X^0_4-\dfrac{3}{2}X^0_7.
  \]
 Let
  \begin{equation}\label{local symm g}
    \hat{X}^{0}=\zeta^0(x,y,y')\dfrac{\partial}{\partial y}=\left(\eta^0(x,y)-y'\xi^0(x,y)\right)\dfrac{\partial}{\partial y}
  \end{equation}
  be the symmetry generator of the ODE $y''=0$ in {\textrm{evolutionary form}}. Therefore, $\zeta^0$ has the form
   \begin{equation}\label{eq7}
     \zeta^0(x,y,y')=\alpha_1xy+\alpha_2 x+\alpha_3\dfrac{y^2}{2}+\alpha_4y+\alpha_5- (\alpha_1x^2+\alpha_3\dfrac{xy}{2}+\alpha_6y+\alpha_7x+\alpha_8)y'.
   \end{equation}
   In {\textrm{evolutionary form}}, the eight point symmetries \eqref{symm:Ypp0:generators} of $y''=0$  are given by
   \begin{eqnarray}\label{r11evol}
     \nonumber 
      \hat{X}^{0}_1 &=& \left(xy-x^2y'\right)\dfrac{\partial}{\partial y},\quad \hat{X}^{0}_2=x\dfrac{\partial}{\partial y},\quad
      \hat{X}^{0}_3=\left({y^2-xyy'}\right)\dfrac{\partial}{\partial y} \\
      \hat{X}^{0}_4 &=& y\dfrac{\partial}{\partial y},\quad \hat{X}^{0}_5=\dfrac{\partial}{\partial y},\quad \hat{X}^{0}_6= yy'\dfrac{\partial}{\partial y},\quad \hat{X}^{0}_7=
      xy'\dfrac{\partial}{\partial y},\quad \hat{X}^{0}_8= y'\dfrac{\partial}{\partial y}.
    \end{eqnarray}
   Let
   \begin{equation}\label{ap1}
     \hat{X} =\left(\zeta^0(x,y,y')+\epsilon \zeta^1(x,y,y')\right)\dfrac{\partial}{\partial y}
   \end{equation}
   be the local approximate symmetry generator admitted by the perturbed ODE \eqref{h101} where $\zeta^0$ is given by equation \eqref{eq7}. The determining equation \eqref{eq005}
   requires
   \begin{equation}\label{eq100}
    \zeta^1_{xx}+2y'\zeta^1_{xy}+{y'}^{2}\zeta^1_{yy}=\left(-\alpha_1y-\alpha_2\right)(y')^{-2}+\left(4\alpha_1
    x-\frac{\alpha_3}{2}y-2\alpha_4+3\alpha_7\right)(y')^{-1}+2\alpha_3x+4\alpha_6.
  \end{equation}
  By change of variable $t=y-xy'$, the homogeneous PDE
  \begin{equation}\label{homogzeta}
    \zeta^1_{xx}+2y'\zeta^1_{xy}+{y'}^{2}\zeta^1_{yy}=0
  \end{equation}
  in  $\zeta^1(x,y,y')=u(x,t)$ becomes a PDE
  \[
    u_{xx}=0,
  \]
  which has a solution $u(x,t)=R_1(t)+xR_2(t)$, where $R_1, R_2$ are arbitrary functions of their arguments. Hence, the PDE \eqref{homogzeta} has the solution $\zeta^1_c=R_1(y-xy')+xR_2(y-xy')$. Now, let
  \[
    \zeta^1_p=P(x,y)(y')^{-2}+Q(x,y)(y')^{-1}+R(x,y)
  \]
  be a particular solution for the nonhomogeneous PDE \eqref{eq100}. Substituting this particular solution into the equation \eqref{eq100} yields the following system of PDEs
 \beq\label{}
\barr
P_{xx}=-\alpha_1y-\alpha_2,\quad  Q_{xx}+2P_{xy}=4\alpha_1
    x-\dfrac{\alpha_3}{2}y-2\alpha_4+3\alpha_7,& \\[2ex]
  R_{xx}+2Q_{xy}+P_{yy}= 2\alpha_3x+4\alpha_6, \quad  2R_{xy}+Q_{yy}=0, \quad R_{yy}=0.&
\earr
\eeq
  Solving the above system gives the general solution of \eqref{eq100}
  \begin{multline}\label{eq10002}
    \zeta^1(x,y,y')=R_1(y-xy')+xR_2(y-xy')-\left(\frac{\alpha_1}{2}x^2y+\frac{\alpha_2}{2}x^2\right)(y')^{-2}\\+\left(\alpha_1
    x^3-\frac{\alpha_3}{4}x^2y-\alpha_4x^2+\frac{3\alpha_7}{2}x^2\right)(y')^{-1}+\frac{\alpha_3}{2}x^3+2\alpha_6x^2.
  \end{multline}
  For simplest solution, take $R_1=R_2=0$. Then, $\zeta^1$ becomes
  \begin{equation}\label{zeta1simple}
    \zeta^1(x,y,y')=\left(-\frac{\alpha_1}{2}x^2y-\frac{\alpha_2}{2}x^2\right)(y')^{-2}+\left(\alpha_1
    x^3-\frac{\alpha_3}{4}x^2y-\alpha_4x^2+\frac{3\alpha_7}{2}x^2\right)(y')^{-1}+\frac{\alpha_3}{2}x^3+2\alpha_6x^2.
  \end{equation}
  Now, we find, one by one, all approximate symmetry components $\zeta^1$ corresponding to each symmetry in \eqref{r11evol}.

  For $\hat{X}^{0}_1$, substituting $\alpha_1=1$, and
  $\alpha_i=0,\,i=2,...,8$ into equations \eqref{eq7} and \eqref{zeta1simple}, we obtain $\zeta^0=xy-x^2y'$ and the corresponding $\zeta^1$ is
  $\zeta^1(x,y,y')=-\frac{1}{2}x^2y(y')^{-2}+ x^3(y')^{-1}$. Hence the corresponding first-order approximate symmetry to $\hat{X}^{0}_1$ is given by
  \begin{equation}\label{hatX1}
    \hat{X}_1=\left(xy-x^2y'+\epsilon\left(\dfrac{1}{2}x^2y(y')^{-2}+ x^3(y')^{-1}\right)\right)\dfrac{\partial}{\partial y}.
  \end{equation}
  It was {\textrm{unstable}} as a point symmetry of the ODE \eqref{ypp:0}, but corresponds to a first-order approximate symmetry \eqref{hatX1}.

  For $\hat{X}^{0}_2$, we have $\zeta^0=x$ and the corresponding $\zeta^1$
  is $\zeta^1(x,y,y')=-\frac{1}{2}x^2(y')^{-2}$. So, $\hat{X}^{0}_2$ used to be an {\textrm{unstable}} point symmetry, but in fact corresponds to a first-order approximate symmetry of the
  perturbed ODE \eqref{h101} given by
  \[
    \hat{X}_2=\left(x-\epsilon\left(\dfrac{1}{2}x^2(y')^{-2}\right)\right)\dfrac{\partial}{\partial y}.
      \]
      Similarly, the {\textrm{unstable}} point symmetry $\hat{X}^{0}_3$ of \eqref{ypp:0} becomes a local first-order approximate symmetry of \eqref{h101} given by
     \[
    \hat{X}_3=\left(y^2-xyy'+\epsilon\left(x^3-\dfrac{1}{4}x^2y(y')^{-1}\right)\right)\dfrac{\partial}{\partial y}.
      \]
   $\hat{X}^{0}_4$ and $\hat{X}^{0}_7$ are not approximate point symmetries of the perturbed ODE \eqref{h101}, while a combination $\hat{X}^{0}_4-\frac{2}{3}\hat{X}^{0}_7$ is an
   {\textrm{evolutionary form}} of the approximate point symmetry $X_9$ in \eqref{symm:Ypp0:BGI:point:generators:approx1}. By
   substituting $\alpha_4=1$, $\alpha_7=2/3$, and other $\alpha_i=0$, one gets $\zeta^0=y-\frac{2}{3} xy'$ and $\zeta_1=0$. A transverse linear combination
   $\hat{X}^{0}_4+\frac{3}{2}\hat{X}^{0}_7$ is the {\textrm{evolutionary form}} for the {\textrm{unstable}} point symmetry $X^0_u$ \eqref{z}. Substituting $\alpha_4=1$,
   $\alpha_7=-3/2$, and other $\alpha_i=0$ into equations \eqref{eq7} and \eqref{zeta1simple}, one gets $\zeta^0=y+\frac{3}{2} xy'$ and $\zeta^1=-\frac{13}{4}x^2(y')^{-1}$. The
   first-order approximate symmetry of the perturbed ODE \eqref{h101} corresponding to the transverse direction of $(\hat{X}^{0}_4,\hat{X}^{0}_7$)-space in \eqref{r11evol} is given by
   \[
     \hat{X}_u=\left(y+\dfrac{3}{2} xy'+\epsilon\left(-\dfrac{13}{4}x^2(y')^{-1}\right)\right)\dfrac{\partial}{\partial y}.
   \]
   $\hat{X}^{0}_5$ is a {\textrm{stable}} symmetry as it is, with $\zeta^1=0$. This easily can be seen by substituting $\alpha_5=1$, and other $\alpha_i=0$ into equation \eqref{zeta1simple}. Similarly, one obtains  $\zeta^1=0$
   corresponding to the {\textrm{stable}} symmetries $\hat{X}^{0}_7$ and $\hat{X}^{0}_8$.

   Finally, $\hat{X}^{0}_6$ is an {\textrm{evolutionary form}} of $X^0_6$ in \eqref{symm:Ypp0:generators}, it should be a
   genuine approximate symmetry coming from {\textrm{stable}} symmetries \eqref{symm:Ypp0:BGI:point:generators:approx2}, here $\zeta^1\neq 0$. Substituting $\alpha_6=1$ and other $\alpha_i=0$ into
   equations \eqref{eq7} and \eqref{zeta1simple} gives $\zeta^0=-yy'$ and $\zeta^1=2x^2$. The corresponding approximate symmetry of the perturbed ODE \eqref{h101} is given by
   \[
    \hat{X}_6=\left(-yy'+2\epsilon x^2\right)\dfrac{\partial}{\partial y}.
      \]
      This is exactly the {\textrm{evolutionary form}} of the approximate point symmetry $X_{11}$ in \eqref{symm:Ypp0:BGI:point:generators:approx2}.
In the following example, we will discuss the stability of exact point and local
symmetries of the Boussinesq differential equation \cite{jafari2009new,manoranjan1988soliton}.
\subsection{Second detailed example}\label{sec:6:4}
  Consider a linear ODE
  \begin{equation}\label{unpert Boss}
    y^{(4)}+y''=0
  \end{equation}
and its perturbed version
  \begin{equation}\label{pert Boss}
  y^{(4)}+y''-\epsilon \left(2yy''+2y'^2\right)=0.
  \end{equation}
The latter ODE can be obtained as a time-independent or a traveling wave reduction of the Bousinesq PDE
\begin{equation}\label{BoussPDE}
  u_{tt}-u_{xx}+\epsilon (u^2)_{xx}-u_{xxxx}=0,\quad u=u(x,t),
\end{equation}
that was introduced by Boussinesq in 1871 to describe the propagation of long waves in shallow water \cite{clarksonz1989}.
First, we seek exact point symmetries for \eqref{unpert Boss} and approximate point symmetries for \eqref{pert Boss}. Let
\begin{equation}\label{exact Boss}
  X^0=\xi^0(x,y)\dfrac{\partial}{\partial x}+ \eta^0(x,y) \dfrac{\partial}{\partial y}
\end{equation}
    be an exact point symmetry generator of the ODE \eqref{unpert Boss}. After the prolongation of $X^0$ to the fourth-order and applying the determining equations
    \eqref{det:eq:exact}, one finds
    \begin{equation}\label{}
       \xi^0 = C_6,\quad \eta^0 = C_1y+C_2+C_3x+C_4\sin x+C_5 \cos x.
    \end{equation}
    Consequently, the ODE \eqref{unpert Boss} admits the following point symmetries:
    \begin{equation}\label{point symm Boss}
      X_1^0 = y\dfrac{\partial}{\partial y},\quad X_2^0=\dfrac{\partial}{\partial y},\quad X_3^0= x\dfrac{\partial}{\partial
      y},\quad X_4^0 = \sin x\dfrac{\partial}{\partial y},\quad X_5^0=\cos x\dfrac{\partial}{\partial
y},\quad X_6^0= \dfrac{\partial}{\partial x}.
    \end{equation}
    Now, we proceed to find approximate point symmetries of the perturbed
    ODE \eqref{pert Boss}. Let
    \begin{equation}\label{}
      X = X^0+\epsilon X^1 = \left(\xi^0(x,y)+\epsilon \xi^1(x,y)\right)\dfrac{\partial}{\partial x}+ \left(\eta^0(x,y)+\epsilon \eta^1(x,y)\right) \dfrac{\partial}{\partial y}
    \end{equation}
be the approximate BGI symmetry generator admitted by the perturbed ODE
\eqref{pert Boss}, where $X^0$ is an exact symmetry generator \eqref{exact Boss} of the
unperturbed ODE \eqref{unpert Boss}. The determining equation for approximate
symmetries \eqref{000} yields
   \begin{equation}\label{}
 \eta^1_{xxxx}+\eta^1_{xx}=0,\quad \eta^1_{xy}=0,\quad \eta^1_{yy}=0,\quad \xi^1_{x}=C_2,\quad \xi^1_{y}=0,\quad C_1=C_3=C_4=C_5=0.
   \end{equation}
 The above system has the solution
 \begin{equation}\label{}
    \xi^1(x,y) = C_2x+a_6,\quad  \eta^1(x,y) =  a_1y+a_2+a_3x+a_4\sin x+a_5 \cos x.
 \end{equation}
 Specifically, the perturbed ODE \eqref{pert Boss} admits six trivial symmetries $X_j=\epsilon X^0_j,\, j=1,2,...,6,$ corresponding to the free constants $a_1\ldots, a_6$, where $X^0_j$ are the exact point symmetries \eqref{point symm Boss} of the unperturbed ODE \eqref{unpert Boss}, and two nontrivial approximate point symmetries
   \begin{equation}\label{x7}
   X_{7}= X^0_2+\epsilon x\dfrac{\partial}{\partial x},\quad X_{8}= X^0_6.
   \end{equation}
      It follows that the only two stable point symmetries of \eqref{unpert Boss} are $X^0_2$ and $X^0_6$, and the {\textrm{unstable}} point symmetries are $X^0_1,\,X^0_3,\,X^0_4$, and $X^0_5$.
\medskip
   \\ Now, we seek exact local symmetries admitted by \eqref{unpert Boss} up to second-order, in the form
    \[
       V^0=\varphi^0(x,y,y',y'')\dfrac{\partial}{\partial y}.
    \]
     Applying the determining equation \eqref{eq05}, one gets
 \begin{equation}\label{}
    \left(D^4 \varphi^0+D^{2}\varphi^0\right)\bigg|_{y^{(4)}=-y''}=0.
   \end{equation}
   The above equation splits into system of PDEs. Solving this system gives
   \beq \label{varphi}
   \barr
   &\varphi^0= k_1+k_2y''+k_3y'+k_4x+k_5y+k_6 \sin x+k_7 \cos x+ k_{8}\left(y'\sin x+{y''}\cos x\right)\\[1ex]
   &+k_9\left(y'^2+{y''}^2\right)+  k_{10}\left(\big({y''}^2-y'^2 \big)\cos x+2y'y''\sin x \right)+ k_{11}\left(\big(y'^2+{y''}^2 \big)\sin x+2y'y''\cos x \right)\\[1ex]
   &+ k_{12}\left(y'\big(2y-x+2{y''}\big)-x{y''}^2 \right)+k_{13}\left(\big(2\sin x-x \cos x\big)y''-\big(x \sin x+\cos x\big)y'+2y\sin x \right)\\[1ex]
   & +k_{14}\left(\big(x\sin x+3\cos x\big)y''+\big(2\sin x-x\cos x\big)y'+y\cos x \right)+k_{15} \left({y''}\sin x-y'\cos x \right).
   \earr
   \eeq
   Hence, the ODE \eqref{unpert Boss} admits the following local symmetries
   \begin{eqnarray}\label{r101}
     \nonumber 
      V_1^0 &=& \dfrac{\partial}{\partial y},\quad V_2^0=y''\dfrac{\partial}{\partial y},\quad V_3^0= y'\dfrac{\partial}{\partial
      y}, \quad V_4^0 = x \dfrac{\partial}{\partial y},\quad V_5^0=y \dfrac{\partial}{\partial y},\quad V_6^0= \sin x \dfrac{\partial}{\partial y},\\
      \nonumber
      V_7^0&=& \cos x \dfrac{\partial}{\partial y},\quad V_8^0= \left(y'\sin x+{y''}\cos x \right) \dfrac{\partial}{\partial y},\quad V_9^0= \left(y'^2+{y''}^2\right) \dfrac{\partial}{\partial y},\\
       \nonumber
       V_{10}^0&=& \left(\big({y''}^2-y'^2 \big)\cos x+2y'y''\sin x \right) \dfrac{\partial}{\partial y},\quad  V_{11}^0= \left(\big(y'^2+{y''}^2 \big)\sin x+2y'y''\cos x \right),\\
       \nonumber
       V_{12}^0&=& \left(2y'\big(y+{y''}\big)-x\big(y'+{y''}^2\big) \right) \dfrac{\partial}{\partial y},\\
        \nonumber
        V_{13}^0&=& \left(\big(2\sin x-x\cos x\big)y''-\big(x\sin x+\cos x\big)y'+2y\sin x \right) \dfrac{\partial}{\partial y},\\
        \nonumber
         V_{14}^0&=& \left(\big(x \sin x+3\cos x \big)y''+\big(2\sin x-x\cos x\big)y'+y\cos x \right) \dfrac{\partial}{\partial y},\\
        \nonumber
        V_{15}^0 &=& \left({y''}\sin x-y'\cos x \right) \dfrac{\partial}{\partial y}.
    \end{eqnarray}
    Now, we will find the approximate local symmetries for the perturbed ODE \eqref{pert Boss}. Let
    \begin{equation}\label{}
     V =\left(\varphi^0(x,y,y',y'')+\epsilon \varphi^1(x,y,y',y'')\right)\dfrac{\partial}{\partial y}
   \end{equation}
   be the local approximate symmetry generator admitted by the perturbed ODE \eqref{pert Boss} where $\varphi^0$ is given by equation \eqref{varphi}. Using the determining equation \eqref{eq005}, one obtains
   \beq
   \barr
   &\varphi^1= Q_1(y)+y''Q_2(y)+a_3x+a_4y'+a_5 \sin x+a_6 \cos x+a_{7}\left(y'\big(2y-x+2{y''}\big)-x{y''}^2 \right)\\[1ex]
   &+a_8\left(y'^2+{y''}^2\right)+a_{9}\left(\big(y'^2+{y''}^2 \big)\sin x+2y'y''\cos x \right) + a_{10}\left(\big({y''}^2-y'^2 \big)\cos x+2y'y''\sin x \right) \\[1ex]
   &+a_{11} \left({y''}\sin x-y'\cos x \right)+a_{12}\left(\big(2\sin x-x\cos x\big)y''-\big(x\sin x+\cos x\big)y'+2y\sin x \right)\\[1ex]
   & +a_{13} \left(y'\sin x+{y''}\cos x\right)+a_{14}\left(\big(x\sin x+3\cos x\big)y''+\big(2\sin x-x\cos x\big)y'+y\cos x \right)\\[1ex]
   &-k_1xy'+k_2 \dfrac{4{y''}^2}{3}+k_4\left(2xy''-\dfrac{x^2y'}{2}+\dfrac{5xy}{2}\right),
   \earr
   \eeq
 and $k_i=0, i=5,...,15$. Consequently, the local symmetries $V_i^0$, $i=5,...,15$ \eqref{r101} of the unperturbed ODE \eqref{unpert Boss} are unstable, while $V_1^0$, $V_2^0$, $V_3^0$ and $V_4^0$ in \eqref{r101} are parts of the approximate symmetries of \eqref{pert Boss} given by
 \[
   V_1=V_1^0 -\epsilon xy'\dfrac{\partial}{\partial y},\quad V_2=V_2^0+\epsilon \dfrac{4{y''}^2}{3}\dfrac{\partial}{\partial y},  \quad V_3=V_3^0,\quad V_4=V_4^0+\epsilon \left(2xy''-\dfrac{x^2y'}{2}+\dfrac{5xy}{2}\right)\dfrac{\partial}{\partial y}.
 \]
 Note that, $V_1$ is an evolutionary form of the approximate point symmetry $X_7$ in \eqref{x7}, while $V_4$ is a second-order approximate symmetry of the perturbed ODE \eqref{pert Boss} corresponds to the unstable point symmetry $X_0^3(=V_4^0)$ in \eqref{point symm Boss}.
    \medskip \\ \textbf{Higher-order approximate symmetries corresponding to unstable point and local symmetries of \eqref{unpert Boss}}\\
    Let
    \begin{equation}\label{local symm g Boss}
    \hat{X}^{0}=\zeta^0\dfrac{\partial}{\partial y}
  \end{equation}
  be the {\textrm{evolutionary form}} of the exact point or local symmetry generator of the unperturbed ODE \eqref{unpert Boss}. Here $\zeta^0=\zeta^0(x,y,y')$ for point symmetries \eqref{point symm Boss}, and $\zeta^0=\phi^0(x,y,y',y'')$ for second-order local symmetries \eqref{r101} of the unperturbed ODE \eqref{unpert Boss}.\medskip
  \\Following Theorem \ref{h.o.approx}, for each unstable local symmetry $V_0^5,...,V_0^{15}$ in \eqref{r101} of the ODE \eqref{unpert Boss}, there is a corresponding higher-order approximate symmetry for the perturbed ODE \eqref{pert Boss} of the form
  \[
      \hat{X} =\left(\zeta^0+\epsilon \zeta^1(x,y,y',y'',y''')\right)\dfrac{\partial}{\partial y}.
    \]
   In particular, $ \zeta^0=y$ corresponds to the {\textrm{unstable}} point symmetry $X_0^1$ in \eqref{point symm Boss}. The corresponding $\zeta^1$ is a solution of
   \[
     \left(D^4 \zeta^{1}+D^{2}\zeta^{1}\right)\bigg|_{y^{(4)}=-y''}=2yy''+2(y')^2.
   \]
   The above equation is a nonhomogeneous PDE in $\zeta^{1}$ and it is solvable. A particular solution of this PDE is given by
   \[
     \zeta^{1}(x,y,y',y'',y''')=\left(\dfrac{x^2}{2}+\dfrac{5}{6}\right){y'}^2+\left(\dfrac{x^2y'+3xy+2y''}{2}\right)y'''.
   \]
   One obtains
    \[
      \hat{X}^1 =\left(y+\epsilon \left(\left(\dfrac{x^2}{2}+\dfrac{5}{6}\right){y'}^2+\left(\dfrac{x^2y'+3xy+2y''}{2}\right)y'''\right)\right)\dfrac{\partial}{\partial y}
    \]
    as a third-order approximate symmetry for the perturbed ODE \eqref{pert Boss} corresponds to the {\textrm{unstable}} point symmetry $X_0^1$. Similarly, one can obtain third-order approximate symmetries corresponding to the point symmetries of \eqref{unpert Boss}.

    In the same way, one can find a third-order approximate symmetry corresponding to each local symmetry in \eqref{r101}. Therefore, let
    \begin{equation}\label{}
     \hat{V} =\left(\varphi^0(x,y,y',y'')+\epsilon \hat{\varphi}^1(x,y,y',y'',y''')\right)\dfrac{\partial}{\partial y}
   \end{equation}
    be approximate symmetry generator for the perturbed ODE \eqref{pert Boss} where $\varphi^0$ is given by the equation \eqref{varphi} . From the determining equation \eqref{eq005}, one can find $\hat{\varphi^1}$ corresponds to each local symmetry of \eqref{r101}. For example, consider the unstable local symmetry $V_9^0= \left(y'^2+{y''}^2\right) \dfrac{\partial}{\partial y}$. By substituting  $\varphi^0=y'^2+{y''}^2$ into the determining equation \eqref{eq005}, one obtains
    \begin{equation}\label{}
    \left(D^4 \hat{\varphi}^1+D^{2}\hat{\varphi}^1\right)\bigg|_{y^{(4)}=-y''}=12y{y'''}^2+56y'y''y'''+10{y''}^3-12y{y''}^2-6{y'}^2y''.
   \end{equation}
   The above equation has a particular solution given by
   \begin{equation}\label{}
     \hat{\varphi}^1=-2x{y''}^2y'''+\dfrac{7}{6}{y''}^3+\left(2y-3xy'\right){y''}^2+\dfrac{1}{2}{y'}^2y''-x{y'}^3.
   \end{equation}
   Hence
   \[
      \hat{V_9} =\left(y'^2+{y''}^2+\epsilon \left(-2x{y''}^2y'''+\dfrac{7}{6}{y''}^3+\left(2y-3xy'\right){y''}^2+\dfrac{1}{2}{y'}^2y''-x{y'}^3\right)\right)\dfrac{\partial}{\partial y}
    \]
    is a third-order local approximate symmetry corresponds to the local symmetry $V_9^0$, that was unstable in the class of second-order local symmetries.

\section{Reduction of order and approximately invariant solutions of perturbed differential equations}\label{sec:sols}
In this section we will introduce the approximate integrating factors and approximate first integrals of perturbed differential equations. Also, we will use the higher-order approximate symmetries to find approximate solutions of some perturbed ODEs.
\subsection{Approximate integrating factors using approximate point symmetries}
  A differential function
  \begin{equation}\label{IF0}
    \mu(x,y,y',...,y^{(n-1)};\epsilon)=\mu_0(x,y,y',...,y^{(n-1)})+\epsilon \mu_1(x,y,y',...,y^{(n-1)})
  \end{equation}
  is an \textit{\textrm{approximate integrating factor}} for the perturbed ODE \eqref{s6} if there is a differential function \, $\phi(x,y,y',...,y^{(n-1)};\epsilon)=\phi_0(x,y,y',...,y^{(n-1)})+\epsilon \phi_1(x,y,y',...,y^{(n-1)})$ such that
  \[
    \mu(y^{(n)}-f_0-\epsilon f_1)=D(\phi)=o(\epsilon).
  \]
Finding the integrating factor allows an approximate reduction of the equation \eqref{s6} to an $(n-1)-$order equation
\begin{equation}\label{IF1}
  \phi(x,y,y',...,y^{(n-1)};\epsilon)=\const+o(\epsilon).
\end{equation}
\begin{remark}
  The integrating factor for the perturbed first-order ODE \eqref{s2} with exact symmetry generator \eqref{Y:1st ode} has the form
  \begin{equation}\label{IF_pert ode}
    \mu(x,y;\epsilon)=\dfrac{1}{\eta-\xi(f_0+\epsilon f_1)},
  \end{equation}
  provided that $\eta \neq \xi(f_0+\epsilon f_1)$. If $(\xi,\eta)$ are analytic in $\epsilon$, then
  \[
    \mu(x,y;0)=\mu_0(x,y)
  \]
  is an integrating factor for the unperturbed first-order ODE \eqref{s1}. Moreover, $\mu(x,y;\epsilon)=\mu_0(x,y)+\epsilon \mu_1(x,y)+o(\epsilon)$ with
  \begin{subequations}\label{approx_IF}
     \begin{eqnarray}
    \mu_0(x,y) &=& \dfrac{1}{\eta^0-\xi^0 f_0}, \\
    \mu_1(x,y) &=& {\mu_0^2}\left(\xi^0 f_1+\xi^1 f_0-\eta^1 \right)
  \end{eqnarray}
  \end{subequations}
is an approximate integrating factor for the ODE \eqref{s2} with approximate symmetry generator \eqref{X:1st ode}.
\end{remark}
As noted in Remark \ref{exact_approx_1st ode_remark}, one has $\xi(x,y;\epsilon)=\xi^0(x,y)+\epsilon \xi^1(x,y)+o(\epsilon)$ and $\eta(x,y;\epsilon)=\eta^0(x,y)+\epsilon \eta^1(x,y)+o(\epsilon)$. By substituting these values into equation \eqref{IF_pert ode} and taking the Taylor expansion about $\epsilon=0$, one reaches to the equations \eqref{approx_IF}.
\begin{example}
  The first-order ODE
  \begin{equation}\label{IF12}
    y'=y+\epsilon xy
  \end{equation}
  admits the approximate symmetry generator
  \[
    X=\left(1+\epsilon\right)y \dfrac{\partial}{\partial y}.
  \]
  The approximate integrating factor for \eqref{IF12} has the form
  \[
    \mu(x,y;\epsilon) = \dfrac{1}{y}\left(1-\epsilon\right).
  \]
  Using this integrating factor, one gets
  \begin{eqnarray}
  \nonumber 
    o(\epsilon) &=& \left(\dfrac{1}{y}\left(1-\epsilon\right)\right)\left(y'-y-\epsilon xy\right) \\
    \nonumber
     &=& \dfrac{y'}{y}-1+\epsilon \left(1-x-\dfrac{y'}{y}\right)  \\
     &=& D\left(\ln y -x +\epsilon \left(x-\dfrac{x^2}{2} -\ln y\right)\right)
  \end{eqnarray}
  Hence
  \begin{equation}\label{IF13}
    \ln y -x +\epsilon \left(x-\dfrac{x^2}{2} -\ln y\right)=C+o(\epsilon)
  \end{equation}
  is a family of approximate solution curves for the perturbed ODE \eqref{IF12}. Note that the first two terms of the Taylor expansion in $\epsilon$ of \eqref{IF13} agree with the first two terms of the Taylor expansion in $\epsilon$ of the exact solution
  \[
    y=C_1e^{ \frac{\epsilon x^2}{2}+x}
  \]
   of the ODE \eqref{IF12}.
\end{example}
\subsection{Determining equations for approximate integrating factors}
For one independent variable $x$ and one dependent variable $y$, the \textit{\textrm{Euler operator}} is given by
\begin{equation}\label{euler-lagrange}
  \dfrac{\delta}{\delta y}=\dfrac{\partial}{\partial y}-D\dfrac{\partial}{\partial y'}+D^2 \dfrac{\partial}{\partial y''}-D^3 \dfrac{\partial}{\partial y'''}+\cdots.
\end{equation}
Since the \textit{\textrm{Euler-Lagrange operator}}\eqref{euler-lagrange} annihilates the total derivative for any differential function, then the integrating factors \eqref{IF0} for the perturbed ODE \eqref{s6} can be found from the following equation:
\begin{equation}\label{Euler-integrating}
  \dfrac{\delta}{\delta y}\left(\mu(y^{(n)}-f_0-\epsilon f_1)\right)=0.
\end{equation}
For the perturbed first-order ODE \eqref{s2}, equation \eqref{Euler-integrating} has the form
\[
  (\mu f_0)_y+\epsilon (\mu f_1)_y+\mu_x=0.
\]
Substituting $\mu=\mu(x,y;\epsilon)=\mu_0(x,y)+\epsilon \mu_1(x,y)$ into the above equation and setting to zero the coefficients of $\epsilon^0$, $\epsilon$, we arrive at the following determining equations for $\mu_0$ and $\mu_1$:
\begin{equation}\label{}
   \mu_{0_x}+(\mu_0 f_0)_y = 0,\quad \mu_{1_x}+(\mu_1 f_0)_y+(\mu_0 f_1)_y =0.
\end{equation}
In particular, for the second-order perturbed ODE
\begin{equation}\label{IF18}
  y''=f_0(x,y,y')+\epsilon f_1(x,y,y'),
\end{equation}
the integrating factor $\mu(x,y,y';\epsilon)=\mu_0(x,y,y')+\epsilon \mu_1(x,y,y')$ for the ODE \eqref{IF18} satisfies
\[
 \dfrac{\delta}{\delta y} \left(\mu(y''-f_0-\epsilon f_1 \right)=0.
\]
The above equation is equivalent to
\begin{equation}\label{IF19}
  y''\mu_y-(\mu f_0)_y-\epsilon (\mu f_1)_y-D \left(y''\mu_{y'}-(\mu f_0)_{y'}-\epsilon (\mu f_1)_{y'}\right)+D^2(\mu)=0.
\end{equation}
Finding the total derivatives appearing in equation \eqref{IF19}, one obtains
\begin{eqnarray}
 \nonumber 
  y'\mu_{yy'}+\mu_{xy'}+2\mu_{y}+(\mu f_0)_{y'y'}+\epsilon (\mu f_1)_{y'y'} &=& 0, \\
  \nonumber
  y'^2\mu_{yy}+2y'\mu_{xy}+\mu_{xx}+y'(\mu f_0)_{yy'}+(\mu f_0)_{xy'}+\epsilon y'(\mu f_1)_{yy'}+\epsilon (\mu f_1)_{xy'}-(\mu f_0)_{y}-\epsilon (\mu f_1)_{y} &=& 0.
\end{eqnarray}
Substituting $\mu(x,y,y';\epsilon)=\mu_0(x,y,y')+\epsilon \mu_1(x,y,y')$ into the above equations, we arrive the following theorem.
\begin{theorem}
  The components $\mu_0$, $\mu_1$, of the approximate integrating factor $\mu(x,y,y';\epsilon)=\mu_0(x,y,y')+\epsilon \mu_1(x,y,y')$ for the perturbed second-order ODE \eqref{IF18} satisfy the following equations
  \begin{eqnarray}
    y'\mu_{0_{yy'}}+\mu_{0_{xy'}}+2\mu_{0_{y}}+(\mu_0 f_0)_{y'y'} &=& 0,\label{IF20} \\[2ex]
    y'^2\mu_{0_{yy}}+2y'\mu_{0_{xy}}+\mu_{0_{xx}}+y'(\mu_0 f_0)_{yy'}+(\mu_0 f_0)_{xy'}- (\mu_0 f_0)_{y}&=& 0,\label{IF21} \\[2ex]
     y'\mu_{1_{yy'}}+\mu_{1_{xy'}}+2\mu_{1_{y}}+(\mu_1 f_0)_{y'y'}+(\mu_0 f_1)_{y'y'} &=& 0,\label{IF22}
       \end{eqnarray}
        \beq \label{IF23}
\barr
   y'^2\mu_{1_{yy}}+2y'\mu_{1_{xy}}+\mu_{1_{xx}}+y'(\mu_1 f_0)_{yy'}+(\mu_1 f_0)_{xy'}-(\mu_1 f_0)_{y}-(\mu_0 f_1)_{y}& \\[2ex]
   +y'(\mu_0 f_1)_{yy'}+(\mu_0 f_1)_{xy'} = 0.
\earr
\eeq
\end{theorem}
\begin{example}
Consider the perturbed Boussinesq ODE
   \begin{equation}\label{pert Boss1}
  y^{(4)}+y''-\epsilon \left(2yy''+2y'^2\right)=0.
  \end{equation}
  Equation \eqref{pert Boss1} can be written in the form
  \[
    D^2(y''+y-\epsilon {y}^2)=0.
  \]
  Hence, the Boussinesq ODE \eqref{pert Boss1} reduces to the second-order ODE
  \begin{equation}\label{IF28}
    y''+y-\epsilon {y}^2=C_1x+C_2.
  \end{equation}
  If $C_1=C_2=\epsilon$, then the ODE \eqref{IF28} has the form
  \begin{equation}\label{IF29}
    y''=-y+\epsilon(x+1+y^2).
  \end{equation}
  Note that, exact solutions of the perturbed equation \eqref{pert Boss1} and ODE \eqref{IF29} are not easy to find. In particular, \verb"Maple dsolve" does not give anything. Using the determining equations \eqref{IF20}-\eqref{IF23}, one can easily find that $\mu=y'+\epsilon(y'-1)$ is an approximate integrating factor for the ODE \eqref{IF29}. Multiplying this integrating factor by \eqref{IF29} yields
  \[
    y'y''+yy'+\epsilon(y'y''-y''+yy'-y-(x+1+y^2)y')=o(\epsilon).
  \]
 We have
  \[
     D\left(y'^2+y^2+\epsilon \left(y'^2-2y'+y^2-(2x+2)y-\dfrac{2y^3}{3}\right)\right)=o(\epsilon).
  \]
  And hence the perturbed Boussinesq ODE \eqref{pert Boss1} is reduced to the first-order ODE
  \begin{equation}\label{IF30}
    y'^2+y^2+\epsilon \left(y'^2-2y'+y^2-(2x+2)y-\dfrac{2y^3}{3}\right)=2c^2+o(\epsilon).
  \end{equation}
  Substituting $y(x;\epsilon)=y_0(x)+\epsilon y_1(x)+o(\epsilon)$ into the ODE \eqref{IF30}, leads to the following system of ODEs
  \begin{eqnarray}
    (y_0')^2+y_0^2 &=& 2c^2, \\
    2y_0'y_1'+2y_0y_1+(y_0')^2-2y_0'+y_0^2-(2x+2)y_0-\dfrac{2y_0^3}{3} &=& 0,
  \end{eqnarray}
 with solutions
 \[
   y_0(x)=c\left(\sin x +\cos x\right),\quad y_1(x)=-\dfrac{c^2 }{3}\sin 2x-\dfrac{c}{2}\left(\cos x+\sin x\right)+C_1\left(\cos x-\sin x\right)+x+c^2+1.
 \]
 Therefore,
 \begin{equation}\label{}
   y(x;\epsilon)=c(\sin x +\cos x)+\epsilon \left(-\dfrac{c^2 }{3}\sin 2x-\dfrac{c}{2}(\cos x + \sin x)+C_1(\cos x - \sin x)+x+c^2+1\right)
 \end{equation}
 is an approximate solution for the Boussinesq ODE \eqref{pert Boss1}.
\end{example}
\subsection{Reduction of order under contact and higher-order symmetries}
 The higher-order approximate symmetry generator for an $n^{\rm th}-$order ODE \eqref{s6}
 \begin{equation}\label{s6sec7}
    y^{(n)}=f_0(x,y,y',...,y^{(n-1)})+\epsilon f_1(x,y,y',...,y^{(n-1)})
  \end{equation}
is given by
\begin{equation}\label{}
  \hat{X}=\hat{ X}^0+\epsilon \hat{X}^1=\left(\zeta^0(x,y,y',...,y^{(s)})+\epsilon \zeta^1(x,y,y',...,y^{(\ell)})\right)\dfrac{\partial}{\partial y}.\quad s,\ell\geq 1.
\end{equation}
The differential functions
   \[
     \omega_k(x,y,y',...,y^{(k)};\epsilon)=\omega_k^0(x,y,y',...,y^{(k)})+\epsilon \omega_k^1(x,y,y',...,y^{(k)})+o(\epsilon),\, k=1,...,n,
   \]
   are called approximate differential invariants for the ODE \eqref{s6sec7} if ${\hat{X}}^{(k)}\omega_k(x,y,y',...,y^{(k)};\epsilon)=o(\epsilon).$ Note that $\omega_k^0$ are exact differential invariants for the unperturbed ODE \eqref{s06}. They arise as constant of integrations of the characteristic equations
   \begin{equation}\label{IF31}
     \dfrac{dy}{\zeta^0}=\dfrac{dy'}{\zeta^{0^{(1)}}}=\cdots=\dfrac{dy^{(k)}}{\zeta^{0^{(k)}}}.
   \end{equation}
   Then the differential invariants $\omega_k^1$ are determined from the following equation
   \[
     H(\omega_{k_y}^1,\omega_{k_{y'}}^1,\cdots)=\hat{X}^{1^{(k)}}(\omega_k^0)\bigg|_{y^{(n)}=f_0},
   \]
    where $H$ is a PDE in $\omega_{k}^1$ resulting from the coefficients of $\epsilon$ in
   \[
     -\left(\hat{X}^{0^{(k)}}(\omega_k)\right)\bigg|_{y^{(n)}=f_0+\epsilon f_1}.
   \]
   \begin{example}
     Consider the second-order ODE given by \eqref{ypp1}
     \begin{equation}\label{ypp1sec7}
      y''=\epsilon (y')^{-1}.
      \end{equation}
     This ODE admits an approximate contact symmetry given by
     \[
    \hat{X}=\hat{X^0}+\epsilon \hat{X^1}=\left(x-\epsilon\left(\frac{1}{2}x^2y'^{-2}\right)\right)\dfrac{\partial}{\partial y}.
      \]
      We determine the invariants $\omega(x,y,y';\epsilon)=\omega^0(x,y,y')+\epsilon \omega^1(x,y,y')+o(\epsilon)$ satisfying ${\hat{X}}^{(1)}\omega=o(\epsilon).$ Clearly, one invariants is $x$. Other invariants are determined by first finding $\omega^0$ satisfying
   \[
    {\hat{X^0}}^{(1)} \omega^0=x\omega^0_y+\omega^0_{y'}=0,
   \]
   which has a solution $\omega^0(x,y,y')=xy'-y$. Then, one finds that $\omega^1(x,y,y')= -\dfrac{1}{2}x^2{y'}^{-1}$ satisfies the PDE
   \[
     x\omega^1_y+\omega^1_{y'}=\dfrac{1}{2}x^2{y'}^{-2}.
   \]
 Hence, $\omega(x,y,y';\epsilon)=xy'-y-\dfrac{\epsilon}{2}x^2{y'}^{-1}+o(\epsilon)$ is an approximate invariant for the ODE \eqref{ypp1sec7}. Now, one can find that $D\omega=D\omega^0+\epsilon D\omega^1=o(\epsilon)$. Thus, the ODE \eqref{ypp1sec7} approximately reduces to the first-order ODE
   \begin{equation}\label{IF32}
     xy'-y-\dfrac{\epsilon}{2}x^2{y'}^{-1}=C+o(\epsilon).
   \end{equation}
   \end{example}
   \begin{example}
   We find an approximate solution for the perturbed Boussinesq ODE \eqref{pert Boss1} using third-order approximate symmetries admitted by \eqref{pert Boss1}. The fundamental solution of the  unperturbed equation \eqref{unpert Boss} is
     \begin{equation}\label{IF33}
         y(x)=C_1x+C_2 \sin x+C_3 \cos x+C_4.
     \end{equation}
     The solution \eqref{IF33} is invariant under the group generated by
     \begin{equation}\label{IF34}
       X_1^0-C_1 X_3^0 - C_2 X_4^0 -C_3 X_5^0-C_4 X_2^0 =\left(y-C_1x-C_2 \sin x-C_3 \cos x-C_4\right)\dfrac{\partial}{\partial y},
     \end{equation}
     where $ X_j^0,\,j=1,...,5$ are the point symmetries \eqref{point symm Boss} for the unperturbed ODE \eqref{unpert Boss}. $X_2^0$ is {\textrm{stable}} as a point symmetry, the corresponding approximate symmetry is  $X^2=(1-\epsilon xy')\partial/\partial y$. At the same time, $X_1^0$,\, $X_3^0$,\, $X_4^0$\, and $X_5^0$ are {\textrm{unstable}} as point symmetries. But using Theorem \ref{h.o.approx}, they correspond to third-order approximate symmetries of \eqref{pert Boss1} given by
     \begin{eqnarray}
       X^1 &=& \left(y+\epsilon\left(\left(\dfrac{x^2}{2}+\dfrac{5}{6}\right){y'}^2+\left(\dfrac{x^2y'+3xy+2y''}{2}\right)y'''\right)\right)\dfrac{\partial}{\partial y},  \\
       X^3 &=& \left(x+\epsilon\left(\dfrac{xy+3x^2y'}{2}+2x^2y'''\right)\right) \dfrac{\partial}{\partial y},
       \end{eqnarray}
       \begin{equation}\label{}
         X^4 = \left(\sin x+\epsilon\frac{\left((3x^2-17)y'-6xy+(3x^2-36)y'''\right)\cos x+\left(15y-12xy'-18xy'''\right)\sin x}{6}\right)\dfrac{\partial}{\partial y},
       \end{equation}
       \begin{equation}\label{}
         X^5 = \left(\cos x+\epsilon\frac{\left((17-3x^2)y'+6xy+(36-3x^2)y'''\right)\sin x+\left(15y-12xy'-18xy'''\right)\cos x}{6}\right)\dfrac{\partial}{\partial y}.
       \end{equation}
       The approximately invariant solution under $X^1-C_1X^3-C_2X^4-C_3X^5-C_4X^2$ is given by
       \begin{equation}\label{IF 34}
         y-C_1x-C_2 \sin x-C_3 \cos x-C_4+\epsilon h(x,y,y',y'',y''')=o(\epsilon),
       \end{equation}
       where $h$ is given by
       \begin{multline}\label{}
         h=\left(\dfrac{x^2}{2}+\dfrac{5}{6}\right){y'}^2+\left(\dfrac{x^2y'+3xy+2y''}{2}\right)y'''-C_1\left(\dfrac{xy+3x^2y'}{2}+2x^2y'''\right) \\
         -C_2\left(\frac{\left((3x^2-17)y'-6xy+(3x^2-36)y'''\right)\cos x+\left(15y-12xy'-18xy'''\right)\sin x}{6}\right) \\ -C_3\left(\frac{\left((17-3x^2)y'+6xy+(36-3x^2)y'''\right)\sin x+\left(15y-12xy'-18xy'''\right)\cos x}{6}\right).
       \end{multline}
       Substitute $y(x;\epsilon)=y_0(x)+\epsilon y_1(x)$ into the equation \eqref{IF 34}, and equate the coefficients of $\epsilon^0$,\, $\epsilon^1$, we find $y_0=C_1x+C_2 \sin x+C_3 \cos x+C_4$ and $y_1=-h(x,y_0,{y_0}',{y_0}'',{y_0}''')$. Hence, the approximate solution of the Boussinesq ODE \eqref{pert Boss1} is given by
       \begin{multline}\label{IF35}
         y(x;\epsilon)= C_1x + C_2 \sin x + C_3 \cos x + C_4 + \epsilon \biggl[\left(\frac{7C_1C_3+5C_2C_4}{2}\right)\sin x + \left(\frac{C_1C_2+2C_3C_4}{2}\right)x\sin x \\ + \frac{C_1C_3 }{2}x^2 \sin x +\left(\frac{15C_2^2 + 17C_3^2}{6}\right)\sin^2 x -\frac{C_2C_3 }{3}\sin 2x + \left(\frac{5C_3C_4-7C_1C_2}{2}\right)\cos x \\+ \left(\frac{C_1C_3-2C_2C_4}{2}\right)x\cos x - \frac{C_1C_2 }{2}x^2 \cos x
        +\left(\frac{17C_2^2+15C_3^2}{6}\right)\cos^2 x +C_1^2 x^2-C_1C_4x-\frac{C_1^2}{3}\biggl].
       \end{multline}
       With the initial conditions $y(0)=1, y'(0)=1, y''(0)=-1, y'''(0)=-1$, the unperturbed ODE \eqref{unpert Boss} has a particular solution
       \begin{equation}\label{Bouss particular soln}
         y(x)=\sin x +\cos x.
       \end{equation}
       Using this particular solution, and the following initial conditions
       \begin{equation}\label{ICs2}
          y(0)=1+\dfrac{16 \epsilon}{3},\quad y'(0)=1-\dfrac{2 \epsilon}{3},\quad y''(0)=-1,\quad y'''(0)=-1+\dfrac{8 \epsilon}{3},
       \end{equation}
        one finds $C_1=0$, $C_2=1$, $C_3=1$, and $C_4=0$. Thus the approximate solution \eqref{IF35} of the perturbed ODE \eqref{pert Boss1} reduces to the following particular approximate solution
        \begin{equation}\label{Bouss particular approx}
          y(x;\epsilon)=\sin x +\cos x +\epsilon \left(\frac{16-\sin 2x}{3}\right).
        \end{equation}
        Note that, an exact solution of the perturbed equation \eqref{pert Boss1} with initial conditions \eqref{ICs2} is not easy to find. Now, we find a numerical solution for the perturbed ODE \eqref{pert Boss1} with the initial conditions \eqref{ICs2} using the fourth-order Runge-Kutta method \cite{momoniat2010symmetry}. In Figure \ref{fig_sec7} below, the solid line represents the exact solution  \eqref{Bouss particular soln} of the unperturbed equation \eqref{unpert Boss}, while the long-dashed light green line refers to the graph of the approximate solution \eqref{Bouss particular approx} of the perturbed equation \eqref{pert Boss1}, and the dotted line represents the Runge-Kutta numerical solution of \eqref{pert Boss1}. The numerical solution for \eqref{pert Boss1} was obtained using the \verb"Maple" package \verb"RK4" with standard tolerances.
   \end{example}
   \begin{figure}[H]
	\centering
		\includegraphics[width=\textwidth]{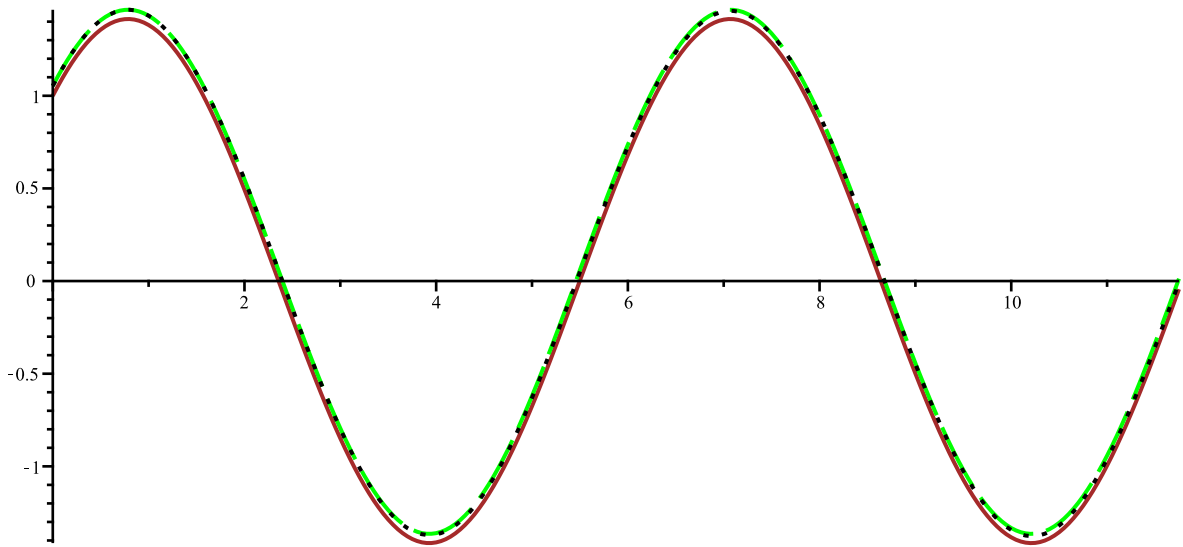}
		\caption{The exact solution \eqref{Bouss particular soln} of the unperturbed equation \eqref{unpert Boss} (dark red solid line) vs. the approximate solution \eqref{Bouss particular approx} of the perturbed equation \eqref{pert Boss1} (light green long-dash line), and the numerical solution of \eqref{pert Boss1} (black dotted line) for \,$\epsilon=0.01,$ with initial conditions \eqref{ICs2}.}\label{fig_sec7}
\end{figure}

\section{Discussion}  \label{conclusion}

In this paper, we considered local symmetries of algebraic and ordinary differential equations involving a small parameter ($\epsilon$), in comparison to the symmetry structure of their unperturbed versions (small parameter equal to zero). We investigated exact symmetries of the unperturbed equations, and exact and approximate symmetries (in the BGI framework \cite{baikov1989, baikov1991, baikov1993}) of the perturbed models. The main goal of the paper was to address the question of stability of symmetries when some given equation is perturbed by addition of small $O(\epsilon)$ terms. It was observed by the original authors of the BGI method that while new and useful approximate symmetries can be sometimes found for perturbed models, some point symmetries of the unperturbed model may not appear in any form in the approximate point symmetry classification of a perturbed model, being thereby \emph{unstable}. The aims of this paper were to find out the conditions under which a symmetry becomes unstable, the form it can assume in the approximate point symmetry classification of a perturbed equation, and applications of approximate symmetries (in particular, higher-order ones) to compute approximate solutions of the given ODE with a small parameter.

For algebraic equations (Section \ref{sec:Alg}), it turned out that every point symmetry of the unperturbed equation is stable: a corresponding approximate symmetry of the perturbed equation always exists. Moreover, approximate symmetry generators of perturbed algebraic equations are more general than the exact symmetry generators of perturbed algebraic equations, since the former contain additional arbitrary functions. An example for a family of exact and perturbed circles in the plane was considered in detail.

For first-order ODEs (Section \ref{sec:1stODEs}), we showed that all exact point symmetries of the unperturbed equations are also stable. Similarly to the case with algebraic equations, for every point symmetry of the unperturbed ODE, there exist families of corresponding exact and approximate symmetries of the perturbed ODE, and the approximate symmetry components arise as first-order Taylor terms in the expansion of exact symmetry components of the perturbed equation in small the parameter.

For second and higher-order ODEs and PDEs, the situation is more complex (Section \ref{sec:higherODEs}): some original symmetries of the unperturbed model \eqref{s06} can be unstable, in the sense of not being inherited as nontrivial approximate point symmetries of a perturbed ODE \eqref{s6} (Example \ref{Ex A perturbed}). At the same time, for some ODEs, all point symmetries of the unperturbed model might be stable (Example \ref{ex:all:stable}). This occurs because in the approximate point symmetry computation of a perturbed ODE, additional conditions on the $O(\epsilon^0)$ approximate symmetry components may or may not arise. The situation is clarified in Section \ref{sec:localsym}, where symmetries (point or local, exact and approximate) are written in the evolutionary form. Theorem \ref{h.o.approx} is proven, showing that to every point or local symmetry of an exact ODE \eqref{s06} of any order, there corresponds an approximate symmetry of the perturbed ODE \eqref{s6}, being possibly a \emph{higher-order symmetry} of order at most $n-1$. Two examples are considered in detail: a nonlinearly perturbed second-order ODE \eqref{h101} (Section \ref{sec:6:3}), and a fourth-order Boussinesq reduction ODE \eqref{pert Boss} (Section \ref{sec:6:4}).

\medskip
In future work, it is important to extend the understanding of relationships between symmetry structures of unperturbed and perturbed models in the cases of systems of ODEs, scalar PDEs, and systems of PDEs. Moreover, it is of high importance to investigate approaches to the computation of approximate symmetry properties of singularly perturbed models, including both ODE models (e.g., Ref.~\cite{o1991singular}) and PDE models, such as almost-inviscid Navier-Stokes fluids and shallow water equations \cite{Wh}.

\section*{Acknowledgments }
A.C. is grateful to NSERC of Canada for research support through a Discovery grant RGPIN-2019-05570.

{\footnotesize
\bibliography{References-19c}
\bibliographystyle{ieeetr}}
\end{document}